\documentclass[useAMS,fleqn,usenatbib]{mnras}
\usepackage[T1]{fontenc}
\usepackage{ae,aecompl}
\usepackage{bethmacros}
\usepackage{graphicx}
\usepackage{hyperref}
\usepackage{amssymb}
\usepackage{amsmath}

\def\spose#1{\hbox to 0pt{#1\hss}}
\def\lta{\mathrel{\spose{\lower 3pt\hbox{$\mathchar"218$}}
     \raise 2.0pt\hbox{$\mathchar"13C$}}}
\def\gta{\mathrel{\spose{\lower 3pt\hbox{$\mathchar"218$}}
     \raise 2.0pt\hbox{$\mathchar"13E$}}}

\title[Entropy Driven Winds]{Entropy Driven Winds: Outflows and Fountains Lifted
Gently by Buoyancy}

\author[Keller et al.]{Benjamin W. Keller$^1$\thanks{Email: benjamin.keller `at'
uni-heidelberg.de},  J. M. Diederik Kruijssen$^1$, James W. Wadsley$^2$
\vspace*{6pt}\\
$^1$Astronomisches Rechen-Institut, Zentrum f{\"u}r Astronomie der Universit\"at
Heidelberg, M{\"o}nchofstra{\ss}e 12-14, D-69120 Heidelberg, Germany\\
$^2$Department of Physics and Astronomy, McMaster University, Hamilton, Ontario,
L8S 4M1, Canada}

\begin{document}
\maketitle
\label{firstpage}
\begin{abstract}
    We present a new theoretical framework for using entropy to understand how
    outflows driven by supernovae are launched from disc galaxies: via
    continuous, buoyant acceleration through the circumgalactic medium (CGM).  When
    young star clusters detonate supernovae in the interstellar medium (ISM) of a
    galaxy, they generate hot, diffuse bubbles that push on the surrounding ISM
    and evaporate that ISM into their interiors.  As these bubbles reach the
    scale height of the ISM, they break out of the disc, rising into the CGM.
    Once these bubbles break out, if they have sufficiently high entropy, they
    will feel an upward acceleration, owing to a local buoyant force.  This
    upward force will accelerate these bubbles, driving them to high
    galactocentric radii, keeping them in the CGM for $>\Gyr$, even if their
    initial velocity is much lower than the local escape velocity.  We derive an
    equation of motion for these entropy-driven winds that connects the ISM
    properties, halo mass, and CGM profile of galaxies to the ultimate evolution
    of feedback-driven winds.  We explore the parameter space of these
    equations, and show how this novel framework can explain both
    self-consistent simulations of star formation and galactic outflows as well
    as the new wealth of observations of CGM kinematics.  We show that these
    entropy-driven winds can produce long wind recycling times, while still
    carrying a significant amount of mass.  Comparisons to simulations and
    observations show entropy-driven winds convincingly explain the kinematics of
    galactic outflows.
\end{abstract}


\section{Introduction}
To zeroth-order, galaxy formation is an accretion process.  Dark matter (DM)
haloes condense from the growth of small initial overdensities, whose distant
counterparts are observable in the cosmic microwave background radiation
\citep{Press1974,Planck2014}.  Baryons trace this structure, accreting onto DM
haloes where they can cool and condense \citep{Spitzer1956,Rees1977,Fall1980},
eventually falling to the centre of the halo where they build up an interstellar
medium (ISM) from which stars may form.  This simple picture, however, presents
us with a number of observational and theoretical problems.  For one,
significantly less than the the expected amount of baryons are found within
galaxies \citep{Papastergis2012}.  Not only are there too few baryons, the
fraction of these baryons that can be found in stars is small $(<30\%)$, and
this fraction has a non-monotonic relationship to the mass of the halo, peaking
for halo masses of $\sim10^{12}\Msun$ \citep{Behroozi2013,Moster2013}.  Despite
the low star formation efficiency seen in galaxies, the metallicity of their
stellar populations and ISM is even lower than what their stellar masses should
have produced \citep{Peeples2014}, while observations of the circumgalactic
medium (CGM) that surrounds these galaxies have found absorption lines for
metals $>100\kpc$ beyond the star forming ISM \citep{Aguirre2001,Martin2002}.
Not only are metals found well beyond the stars that created them, but modern
surveys of the CGM surrounding $L_*$ galaxies have found ubiquitous cool
$T\sim10^4\K$ gas \citep{Morganti2003,Steidel2010}, which cannot be in
pressure equilibrium with the volume filling hot $T\sim10^6\K$ CGM
\citep{Putman2012}.  All of these problems point towards a single potential
solution: galaxies not only accrete gas, they eject it as well.  Outflows driven
by stellar feedback \citep{Larson1974,Keller2015}, active galactic nuclei (AGN)
\citep{DiMatteo2005,Keller2016}, radiation \citep{Murray2011}, or cosmic rays
\citep{Ipavich1975} can all act to remove gas from the disc of a galaxy,
dropping the baryon fraction, starving the star formation process, and polluting
the CGM with metals and cold, entrained gas.  How these outflows are launched,
and the trajectory they take through the galactic corona is one of the most
important unsolved problems in modern galaxy formation theory.  In this paper,
we present a novel framework that helps explain not only how stellar feedback
can launch galactic outflows, but why these outflows are so effective at
removing mass and metals from the ISM of galaxies at or below the mass of our
own.

\subsection{Direct and Indirect Evidence for Galactic Outflows}
The retention fraction of a galaxy's baryon budget (which totals $M_{bary} = f_BM_{vir}
\sim 0.16M_{vir}$, \citealt{Planck2014}) varies for
galaxies in different mass regimes, but for $L_*$ galaxies like our own, with
$M_{vir}\sim10^{12}\Msun$, typically $5-30\%$ of the baryon budget is found in
stars, and a comparable amount within the disc ISM.  This leaves roughly half of
the baryons unaccounted for by the components in the galactic disc, the
so-called ``missing baryon'' problem
\citep{Fukugita1998,Cen1999,Papastergis2012}.  Three possible solutions are that
the remaining baryons are either residing within the CGM, have been expelled
from the galaxy altogether, or never accreted onto the disc to begin
with \citep{White1978}.  Evidence from simulations points towards most of the
baryons accreted by a halo remaining in the CGM, rather than being ejected much
beyond the virial radius \citep{Muratov2015,Christensen2016,Keller2016}.  Recent
high signal-to-noise observations of the X-ray absorption due to {\sc Ovii}
suggest that the majority of these cosmological ``missing'' baryons are indeed
in a warm/hot medium with $T\sim10^6\K$ surrounding galaxy overdensities
\citep{Gupta2012,Nicastro2018}.

An even clearer sign of significant mass transport by galactic outflows is the
presence of metals in the CGM and IGM \citep{Songaila1996,Aguirre2001}, combined
with a relative lack of metals within the ISM and stars of galactic discs
\citep{Bouche2007,Peeples2011,Peeples2014}.  Unlike the simple deficiency of
baryons that census studies like \citet{Fukugita1998,Papastergis2012} detect,
so-called preventive feedback \citep{White1978} cannot explain this missing
metal problem, as these metals were not accreted on to the disc, but formed
within it.  Only by ejecting them through galactic outflows can these
observations be explained \citep{Dave1998,Finlator2008,Shen2010}.

A far less simple sign of (but perhaps more subtle diagnostic for) galactic
outflows is the relation between the stellar content of a galaxy and its halo
mass.  From the free-fall time of gas in spiral galaxies, we would expect
that the vast majority of baryons would have formed stars within $<<t_{hubble}$,
and yet we find no more than $\sim5\%$ of the total mass of galaxies in stars in even
the most massive haloes \citep{Behroozi2013,Moster2013}.  A low star formation
efficiency may be explained by purely intragalactic processes, such as the
disruption of molecular clouds \citep{Murray2010,Walch2012} or the injection of
turbulence \citep{Larson1981,Joung2006}.  They may also be explained by purely
extragalactic processes, such as inefficient cooling of the CGM
\citep{Rees1977}.  However, most three-dimensional hydrodynamic simulations of
galaxy formation find that galaxies both above
\citep{Scannapieco2012,Munshi2013} and below \citep{Governato2004,Pontzen2012}
$L_*$ suffer from catastrophic overcooling and runaway star formation without
some form of strong, ejective feedback powering galactic outflows and winds
\citep{Hopkins2012b,Keller2015}.  The ubiquitous failure of $L_*$ galaxy
simulations to regulate their star formation revealed by
\citet{Scannapieco2012}, combined with the abundance matching evidence for
baryon reduction from \citet{Behroozi2013} and \citet{Moster2013} made clear
that the ability for stellar feedback to drive outflows is critical to galaxy
formation.

\subsection{Galactic Outflows in Simulations and Semi-Analytic Modelling}
With all of the previous evidence for the important role outflows play, it is
unsurprising that feedback-driven outflows are important components of both
semi-analytic models \citep{Somerville1999,Bertone2007} as well as hydrodynamic
simulations \citep{Springel2003,Dallavecchia2008,Oppenheimer2008} of galaxy
evolution.  Until recently, both of these methods required phenomenological
descriptions of how outflows actually were launched from the ISM and travelled
through the CGM.  The most simple (and common) of these models assume a simple
functional form for the mass outflow rates, and velocities
\citep{Springel2003,Bertone2007,Oppenheimer2008}.  These are often coupled to
each other through a mass loading factor $\eta$, such that for a given outflow
velocity $v_{wind}$, $\frac{1}{2}\eta v_{wind}^2 \sim 10^{49}\erg/\Msun$,
roughly the specific energy injection available due to supernovae
\citep{Leitherer1999}.  In order to vary the effectiveness of these outflows in
different halo scales, the functional form of $\eta$ and $v_{wind}$ often take
into account the halo mass \citep{Bertone2007}, the gravitational potential
\citep{Oppenheimer2008}, or even the stellar mass \citep{Dave2016}.   Typically,
fairly large $(\gg 200\kms)$ values are chosen for the outflow velocities $v_{wind}$, in order
to ensure that the winds can actually remove material from the ISM (or in the
case where $v_{wind} > v_{esc}$, from the halo entirely, e.g.
\citealt{Bertone2007}).  The effectiveness of winds are often furthermore
enhanced by decoupling them from the hydrodynamic solver
\citep{Springel2003,Oppenheimer2008,Dave2016}, thereby artificially suppressing
drag (which may be numerically enhanced due to poor resolution).  Thus, these
outflows evolve {\it ballistically}: with a large initial velocity, imparted
directly while in the ISM, followed by an orbit through the halo unimpeded by
drag (until hydrodynamic forces are re-enabled).  

It has now become possible, with the advent of better numerical techniques and
more powerful computing hardware to simulate galactic winds in a more
self-consistent fashion.  Much numerical work has gone into avoiding issues that
occur when feedback is deposited into regions with insufficient resolution,
losing its thermal energy to numerical overcooling
\citep{DallaVecchia2012,Hopkins2014,Keller2014}.  These models allow for gas to
be directly heated by SN, and subsequently accelerated due to either
hydrodynamic forces \citep{DallaVecchia2012,Keller2014}, or due to momentum
imparted by the SN ejecta \citep{Agertz2013,Hopkins2014}.  The winds driven from
these galaxies have been found to roughly agree with observed properties
\citep{Keller2015,Muratov2015,vandeVoort2016}, and appear to be successful (in
some cases at least) in producing realistic disc galaxies
\citep{Hopkins2014,Crain2015,Keller2016}.  Unlike the simple ballistically
driven, phenomenological models of galactic winds, the origin of scaling
relations observed in these self-consistent models is not obvious.  Wind models
with imposed scalings also feature multiple free parameters, some of which may
even be non-local (such as halo mass).  This makes for a large volume of
parameter space which must be explored through trial and error to stumble upon
parameters that can yield realistic galaxies.  In general, high mass loadings
are seen in smaller, high redshift objects \citep{Keller2015,Muratov2015}, but
how this relationship arises is unclear.  High resolution simulations of wind
launching sites rarely see sufficiently high velocities to escape the galactic
halo, or even rise to particularly large radii \citep{Girichidis2016b,Li2017b}.
Indeed, most cosmological simulations rarely see outflows travelling fast enough
to leave the halo \citep{Muratov2015,Christensen2016,vandeVoort2016}, yet they
are seen to persist in the CGM for much longer than would be expected from
simple ballistic recycling \citep{Christensen2016}.


\subsection{Structure and Kinematics of the CGM}
Recent observations have begun to reveal that the CGM is a
multiphase, non-equilibrium environment \citep{Stocke2013,Werk2014}.  One of the
most important of these studies is the COS-Halos survey \citep{Werk2014}.  By observing
UV absorption in the haloes of 44 galaxies, from $L\sim0.1L_*-3L_*$ out $160\kpc$
at fairly low redshift ($z\sim0.2$), COS-Halos has provided constraints on the
temperature, density, metallicity, and kinematic state of the typical $L_*$ CGM.
COS-Halos has revealed that there is a ubiquitous, multiphase medium surrounding
galaxies with a density profile for cool gas of $n_H\propto(R/R_{vir})^{-0.8\pm
0.3}$ \citep{Werk2014}, containing a mixture of cool ($T\sim10^4\K$) and
warm ($T>10^5\K$) gas that is out of hydrostatic equilibrium
\citep{Werk2014,Werk2016}, and should rain out of the CGM on relatively short
timescales.This material accounts for {\it at least} half of
the missing baryons from $L_*$ haloes \citep{Werk2014,Prochaska2017}, and similar
fractions for the missing metals \citep{Peeples2014,Prochaska2017}.  One of the
most puzzling results from COS-Halos is that despite the failure of single or
multiphase hydrostatic models to fit the observed density and velocity profiles
\citep{Werk2014}, the velocities of the warm component is fairly slow, with
$\Delta v \sim 100-200\kms$ \citep{Werk2016}, well below the escape velocities
of the systems observed.  For one, this warm material should be very fast if in
fact the cool gas (which has similar velocities to this, see \citealt{Stocke2013}) has
been entrained in a hot wind.  For another matter, if this material is moving
well below the escape velocity, it should re-accrete on very short timescales,
implying that the galactic winds are ineffective at removing material from the
ISM of galaxies.  Reconciling this with the low metal content and star formation
efficiency seen in observations, together with the results of numerical
simulations is a major problem for ballistically evolving, energy or momentum
driven winds.

If we are to understand the evolution of outflows, it is important to not just
understand the mechanism that powers these outflows, but the environment through
which they travel.  We now know that the CGM contains a significant amount of
material, and that outflows ejected from the galaxy must interact with this
material as they travel.  For a given Lagrangian parcel of outflowing fluid, it
will spend the majority of its life cycle in contact with the hot, volume-filling
component of the CGM.  It is therefore important to know what the volume
filling, hot phase of the CGM looks like in order to calculate the effects of
two critical, often ignored effects: buoyancy and drag.  Buoyancy is the force
which drives the convective instability, and occurs when a parcel of fluid sees
an entropy gradient $\nabla S$ which points in the opposite direction to the
gravitational gradient $\nabla \phi$ \citep{Chandrasekhar1961}, and depends
primarily on the entropy profile of the CGM.  Drag is the loss of momentum to
the surrounding fluid, and depends on both the density profile of the CGM and
the cross-sectional area and velocity of the outflowing fluid.  In the limit of
a perfect vacuum surrounding galaxies, these effects vanish, and outflows evolve
ballistically.  However, COS-Halos and other studies
\citep{Steidel2010,Gupta2012,Putman2012,Stocke2013} reveal that the CGM contains
significant mass, and this mass will interact with outflowing material, either
imparting it with (through buoyancy or gravity) or robbing it of momentum (through drag).  

The entropy profiles of massive $>10^{13}\Msun$ haloes have been well-studied, as
their virial temperatures are large enough to make them detectable (and
mappable) through their x-ray emission \citep{Spitzer1956,Cowie1980,Kaiser1991}.
X-ray observations allows the mapping of both both the surface density and
temperature of their CGM.  These studies suggest that entropy profile can be fit
by a simple power law.  The index of this power law was first predicted by
\citet{Kaiser1991}, with an index of $\alpha\sim1.1$. Observations have
confirmed the rough picture of a power law $K\propto r^\alpha$ with a
$\alpha\sim0.9-1.5$ slope \citep{Werner2012,Babyk2018}.  Meanwhile, the density
of the volume-filling phase has been mostly found to follow a power law
$\rho\propto^{-\beta}$ with slope $\beta\sim1-2$ \citep{Werk2016,Bregman2018},
which, as we show in section 2.4 of this paper, is consistent with a hot halo
with entropy slope of $\alpha\sim1.1$ in hydrostatic equilibrium.  Within these
massive haloes, phenomena such as cooling flows \citep{Cowie1980}, buoyant AGN
bubbles \citep{Sanders2005}, and the formation of stable entropy profiles and
cores \citep{Babyk2018} have all been observed.  In these more massive systems, cooling times tend to
be relatively long ($\sim\Gyr$), while Mach numbers are small, allowing much of the
halo evolution to be simplified to adiabatic, incompressible hydrodynamic
interactions.  The evolution of overdensities within this medium has been of
primary concern.  These overdensities can be subject to thermal
\citep{Nulsen1986,Balbus1989} and convective
\citep{Loewenstein1989,McNamara2007} instabilities, driving material down to the
center of the halo where it can fuel star formation and fuel AGN.  These AGN can
then in turn heat the halo gas, driving either suppressing these instabilities
\citep{Nulsen1986,McNamara2007} or driving them towards a self-regulated
equilibrium \citep{Voit2017}.  The recent work by \citet{Voit2017} nicely
summarizes much of this theoretical work, and develops an analytic framework for
understanding the interaction between AGN-driven outflows and convective/thermal
instabilities in the CGM.

\subsection{Entropy Driven Winds: Connecting the ISM, CGM, and Outflows} 
In simulations that allow winds, fountains, and outflows to be launched
self-consistenty \citep{Muratov2015,Keller2016,Agertz2013}, understanding the
behaviour of these outflows, and how they scale with various galaxy properties
is difficult.  It is clear that the state of the ISM as well as the properties
of the galaxy halo and hot CGM all play a role in setting the mass loadings,
recycling times, and velocities of outflows.  In particular, we have found in
\citet{Keller2015,Keller2016} that the effectiveness of supernovae-driven winds
falls precipitously once the halo mass exceeds $10^{12}\Msun$.  Understanding
this behaviour requires a theory that takes into account gravity as well as
hydrodynamic effects from the ISM and CGM, that can explain the typical
observed kinematics \citep{Werk2014}, as well as the relatively long
re-accretion times for mass-loaded winds seen in simulations
\citep{Keller2015,Christensen2016,Muratov2015}.

In this paper, we present a new model for understanding the flows of material
out of the galaxy.  Building on previous studies that developed the theory of
convective and thermal instability in the haloes of massive galaxies, as well as
analytic models for superbubbles driven by clustered supernovae, we show how
buoyancy can drive outflows to high galactic radii in Milky Way like $L_*$
galaxies.  Accelerated gently through the CGM, rather than ballistically from
within the ISM, entropy driven fountains can have recycling times of $>\Gyr$
without reducing their mass loading.  We also examine how buoyancy and radiative
cooling sets important limits on the effectiveness of supernovae feedback in
haloes more massive than $10^{12}\Msun$, and when the gaseous disc is dense and
thin.  Finally, we compare the behaviour of entropy-driven outflows to
cosmological galaxy and high-resolution ISM simulations, as well as observations
of gas flows within the CGM.

\begin{figure}
    \includegraphics[width=0.5\textwidth]{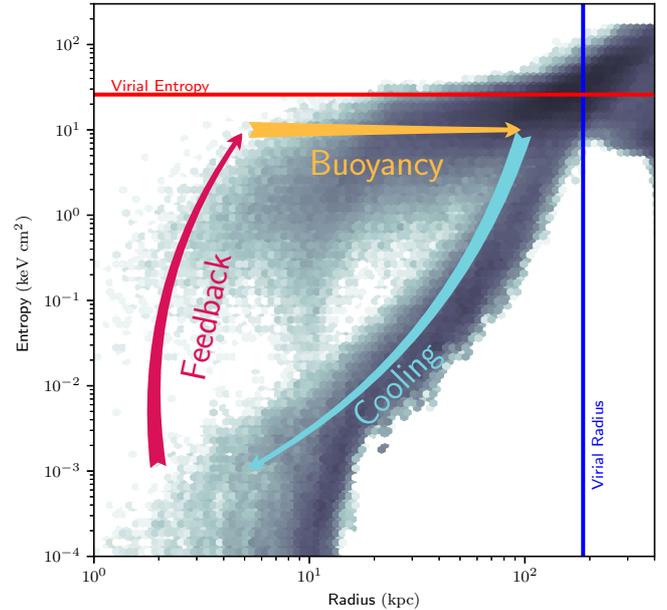}
    \caption{The flow of gas through the CGM in the entropy-driven wind
    framework.  Feedback drives gas to high entropies within the ISM, which
    then break out of the disc and rise buoyantly, conserving their entropy as
    they expand adiabatically.  Eventually, the temperature of feedback heated
    gas drops to the peak of the cooling curve at $\sim10^5$K.  The gas can then
    lose entropy, and thus become negatively buoyant.  This allows it to accrete
    back upon the disc, and potentially continue the cycle once more.  The
    background histogram shows gas particles from the MUGS2 galaxy g1536.}
    \label{entropy_story}
\end{figure}

\section{Deriving the Evolution of Entropy Driven Winds}
\subsection{CGM Entropy}
In most studies of hot coronae around galaxies \citep{Cowie1980,Voit2017}, the
thermodynamic entropy $S$ is replaced with the more easily-measured and
expressed adiabatic invariant $K$, defined simply using the temperature $T$ and
number density $n$ of a parcel of fluid:
\begin{equation}
	K = k_BTn^{1-\gamma}
	\label{K}
\end{equation}
This is related to the specific Clausius entropy $S$ (for a fluid with specific
isochoric heat capacity $c_V$) simply as:
\begin{equation}
	S = c_V \ln K
	\label{entropy}
\end{equation}
While, formally, $K$ is not the true thermodynamic entropy, we will follow the
convention of referring to this as ``entropy'' for the remainder of this paper.

From this simple definition, we can derive for a halo a ``virial entropy'',
the entropy of a virialized halo at the virial radius $R_{200}$.  If the virial
radius is defined as the radius within which a $M_{200}$ halo has average density
$\rho_{200} = 200\rho_{crit}$:
\begin{equation}
	R_{200} =
	\left(\frac{4}{3}\pi(\rho_{200})\right)^{-1/3}M_{200}^{1/3}
	\label{virial_radius}
\end{equation}
Gas at this radius has a characteristic virial temperature $T_{200}$, such that
the gas has thermal/kinetic energy equal to half the gravitational potential
energy of the halo:
\begin{equation}
	T_{200} = \frac{2GM_{200}}{3R_{200}}\frac{\mu m_p}{k_B}
	\label{virial_temp}
\end{equation}
Together, equations~\ref{virial_radius} and~\ref{virial_temp} can be used to
derive a virial temperature for a halo of mass $M_{200}$:
\begin{equation}
	T_{200} = \frac{G\mu m_p}{k_B}\left(\frac{
	4\pi\rho_{200}M_{200}^2}{27}\right)^{1/3}
	\label{virial_temp2}
\end{equation}
Therefore, when we define using equation~\ref{K} our virial entropy $K_{200}$
\begin{equation}
	K_{200} = k_B T_{200}\left(\frac{f_B\rho_{200}}{\mu m_p}\right)^{1-\gamma}
	\label{virial_entropy}
\end{equation}
where $f_B$ is the baryonic mass fraction of the halo.  We shall use a value of
$f_B=0.16$ for this study.
We can use equation~\ref{virial_temp2} to derive an expression for the virial
entropy for a halo of mass $M_{200}$, given an adiabatic index of $\gamma=5/3$:
\begin{equation}
	K_{200} = G\left(\frac{f_B^2\pi\mu^5 m_p^5
	M_{200}^2}{150\rho_{crit}}\right)^{1/3}
	\label{virial_entropy2}
\end{equation}
In units of ${\rm keV\;cm}^2$, this gives the simple relation for ionized gas with solar
metallicity (and thus a mean molecular weight of $\mu=0.62$):
\begin{equation}
	K_{200} = 30.06\;{\rm keV\;cm}^2
	\left(\frac{M_{200}}{10^{12}\Msun}\right)^{2/3}
	\label{virial_entropy_phys}
\end{equation}
For the rest of this paper, we will continue using an adiabatic index of
$\gamma=5/3$ and a mean molecular weight of $\mu=0.62$.

\subsection{Superbubble Entropy}
We can also estimate the entropy generated in feedback-heated superbubbles.
Following calculations from \citet{Weaver1977} and \citet{MacLow1988}, we can
calculate the interior density of the superbubble at time $t_7 =
t/10\Myr$, driven by a luminosity of $L_{38} = L/10^{38}{\rm erg\;s}^{-1}$ in an
ambient medium of density $n_{amb} (\hcc)$, as a function of the distance $r$ from the
centre of a superbubble of radius $R_{SB}$:
\begin{equation}
    n(r) = 4\times10^{-3}\hcc\; L_{38}^{6/35} n_{amb}^{19/35}
    t_7^{-22/35}(1-r/R_{SB})^{-2/5}
    \label{SB_density}
\end{equation}
Integrating this equation from $0$ to $R_{SB}$, we find the total mass contained
within the bubble is:
\begin{equation}
    m_{SB} = \frac{125}{39}\pi\mu m_pn(0)R_{SB}^3
    \label{SB_mass}
\end{equation}
In the limit of weak cooling, the partition of energy in a luminosity-driven
bubble gives the internal energy of $U=5/11Lt$.  The mean temperature of the
bubble interior is therefore given as:
\begin{equation}
    T = \frac{26}{275k_B\pi}n(0)^{-1}R_{SB}^{-3}Lt
    \label{SB_Tmean}
\end{equation}
We assume during the superbubble break out process, the interior becomes
well-mixed. This allows us to apply the \citet{Kompaneets1960} approximation to
the hot bubble's density as well as pressure, and this equation and the average
density of the superbubble to determine the entropy of outflowing material using
equation~\ref{K}:
\begin{equation}
    K_{SB} =
    \frac{26}{275\pi}\left(\frac{375}{156}\right)^{-2/3}n(0)^{-5/3}R_{SB}^{-3}Lt
    \label{SB_entropy_time}
\end{equation}

\begin{figure*}
    \includegraphics[width=\textwidth]{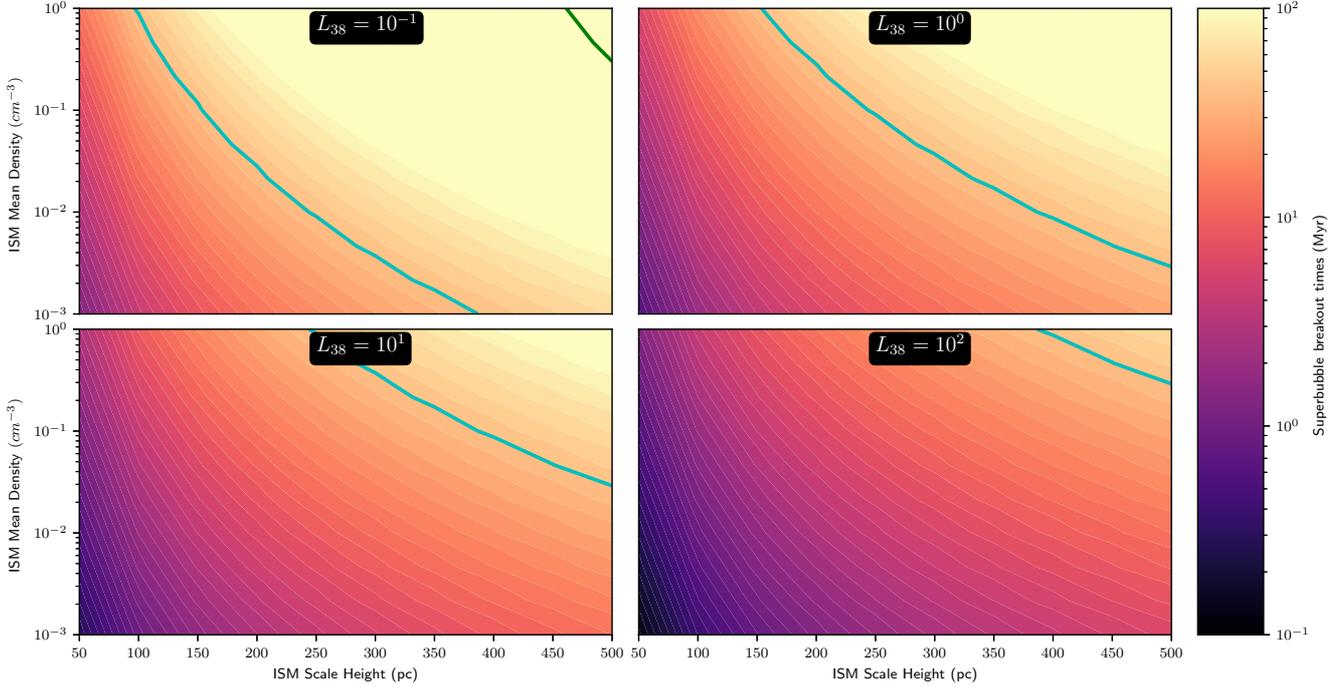}
    \caption{The break out time of  a superbubble driven by four different mass
    star clusters, in a range of different ISM environments (mean density vs.
    scale height).  The four panels show the break out time for a star cluster of
    mass $10^3\Msun$ (upper left), $10^4\Msun$ (upper right), $10^5\Msun$ (lower
    left), and $10^6\Msun$ (lower right).  The cyan curves show break out times
    of $40\Myr$.  Regions to the right of this curve cannot break out of the
    disc before their SN shutoff, breaking the assumption of constant-luminosity
    driving.  The green curve shows $0.1t_{cool}$, the time required for the
    interior of the superbubble to lose $10\%$ of its thermal energy.  As is
    clear here, even the smallest star clusters can drive a superbubble out of
    the ISM before an appreciable amount of energy is lost to cooling.}
    \label{SB_breakout_times}
\end{figure*}

The break out of the superbubble-heated gas occurs when the radius of the bubble
is roughly equal to the scale height of the ISM $R_{SB}\sim h$
\citep{Kruijssen2019b}, so that we can take the equation for the bubble radius
$R_{SB}$:
\begin{equation}
    R_{SB} = 267\pc\; L_{38}^{1/5} n_{amb}^{-1/5} t_7^{3/5}
    \label{SB_R}
\end{equation}
and invert it to solve for $t_7$ with $R_{SB} = h$:
\begin{equation}
    t_7 = \left(\frac{h}{267\pc}\right)^{5/3} L_{38}^{-1/3} n_{amb}^{1/3} 
    \label{SB_breakout_time}
\end{equation}
This gives a superbubble density at breakout that scales as $n_{SB}\propto
L^{8/21}h^{-22/21}n^{1/3}$.  If we assume the stellar cluster driving this
superbubble is a single stellar population with a \citet{Chabrier2003}-like IMF,
then $L_{38}\sim M_{cluster}/10^4\Msun$.  Recent observations show typical young
stellar clusters (YSCs) have masses in the range of $10^4-10^6\Msun$
\citep{Longmore2014,Ginsburg2018,Mok2019}. The last SNe detonates $\sim 30\Myr$ after the
birth of the cluster, and thus the assumption of a constant luminosity driving
is correct so long as $t_7 < 3$.  As is shown in figure~\ref{SB_breakout_times},
this is the case for a large fraction of the ISM/star cluster configuration
space.

This  can finally be combined with equation~\ref{SB_entropy_time} to derive the
entropy of superbubble gas as it breaks out from the ISM of a disc with
scale height $h$.
\begin{equation}
    K_{SB} = 5.84\;{\rm keV\;cm}^2\left(\frac{h}{267\pc}\right)^{26/63}
    L_{38}^{2/63}
    n_{amb}^{-14/63}
    \label{SB_entropy}
\end{equation}
If this hot bubble has entropy greater than the inner corona, it will experience
a buoyant force upwards.

We may also calculate a maximum entropy in the case of a poorly-mixed bubble,
where the most buoyant fraction of the superbubble will accelerate away from the
lower-entropy edges of the bubble.  In this case, we can use the equation for
the peak temperature, together with equation ~\ref{SB_density}:
\begin{equation}
    T(r) = 3.5\times 10^6\K\;  L_{38}^{8/35} n_{amb}^{2/35} t_7^{-6/35}(1-r/R_{SB})^{2/5}
    \label{SB_temperature}
\end{equation}
Setting $r=0$ in both of these equations allows us to derive a maximum
superbubble entropy $K_{max}$:
\begin{equation}
    K_{max} = 11.97\;{\rm keV\;cm}^2\; L_{38}^{4/35} n_{amb}^{-34/105} t_7^{26/105}
    \label{SB_max_entropy_time}
\end{equation}
Which, together with equation~\ref{SB_breakout_time}, can derive a final value
for this maximum entropy:
\begin{equation}
    K_{max} = 11.97\;{\rm keV\;cm}^2\left(\frac{h}{267\pc}\right)^{26/63}
    L_{38}^{2/63}
    n_{amb}^{-76/315}
    \label{SB_max_entropy}
\end{equation}
As can be seen here, the maximum entropy is $\approx 2$ times the average
entropy in a superbubble venting from a typical disc galaxy.  Note however the
slightly stronger density dependence on the maximum entropy.  Naturally, this
high-entropy component will carry substantially less mass out of the galaxy, and
is therefore unlikely to be an effective measure of the entropy in the heavily
mass-loaded winds predicted by theory, simulations and observed in absorption
line studies.

\subsection{Equations of Motion for a Buoyant Bubble}
Beginning with the Euler equation for momentum conservation, we can derive an
equation of motion for a buoyant bubble rising through a medium in a
gravitational potential $\phi$:
\begin{equation}
    \frac{D\mathbf{u}}{Dt} = \frac{\nabla P_{SB}}{\rho_{SB}} - \nabla\phi
    \label{euler_momentum}
\end{equation}
If the sound crossing time inside the superbubble is less than the time it takes
for the superbubble to cross a pressure scale length for the background CGM
\citep{Kompaneets1960}, then it will stay in pressure equilibrium with the
environment, such that $P_{SB}\sim P$.  As the density of the superbubble can be
expressed in terms of pressure and entropy using equation~\ref{K} ($\rho = \mu
m_p (P/K)^{3/5}$), we can re-write the momentum equations as:
\begin{equation}
    \frac{D\mathbf{u}}{Dt} = \frac{\nabla P}{\mu m_p P^{3/5}}K_{SB}^{3/5} -
    \nabla\phi = \frac{\nabla P}{\rho}\left(\frac{K_{SB}}{K}\right)^{3/5} -
    \nabla\phi
    \label{momentum_entropy}
\end{equation}
If the background CGM is in hydrostatic equilibrium, then $\nabla P/\rho =
\nabla\phi$, and we can write an equation for the Lagrangian acceleration of the
hot bubble through an entropy-stratified, hydrostatic CGM (Archimedes'
principle, in entropy form):
\begin{equation}
    \ddot{r} =
    \nabla\phi\left[\left(\frac{K_{SB}}{K(r)}\right)^{3/5}-1\right]
    \label{acceleration_phys}
\end{equation}

The first, positive term of this equation is the buoyant force due to the local
entropy gradient, while the second, negative term is simply the gravitational
acceleration. As is clear from this, when the superbubble entropy $K_{SB}$ is
greater than the CGM entropy $K(r)$, the bubble will feel a positive, upward
buoyant acceleration.  In the limit where $K_{SB}$ goes to zero, it will
experience gravitational free-fall.  If our CGM entropy profile is set by a
simple power-law:
\begin{equation}
    K(r) = K_{200}\left(\frac{r}{R_{200}}\right)^\alpha
    \label{CGM_entropy}
\end{equation}
We can then derive an acceleration for our superbubble in terms of the virial
entropy $K_{200}$ and the bubble's entropy $K_{SB}$ and height in the halo $r$,
which can be simplified using the assumption of a flat rotation curve with
$\nabla\phi=v_c^2/r$:
\begin{equation}
    \ddot{r} =
    \frac{v_c^2}{r}\left[\left(\frac{K_{SB}R_{200}^\alpha}{K_{200}r^\alpha}\right)^{3/5}-1\right]
    \label{acceleration_phys_nodrag}
\end{equation}

Setting the left-hand side of this equation to zero tells us that, as we might expect,
the bubble ceases to be buoyant when it reaches a radius where its entropy is
equal to the surrounding material (which can also be derived directly from the
power-law entropy profile):
\begin{equation}
    r_{max} = \left(\frac{K_{SB}}{K_{200}}\right)^{1/\alpha}R_{200}
    \label{maximum_radius}
\end{equation}

\subsection{Damping by Drag}
When we include the effects of drag, we can derive the final set of equations
governing the motion of the bubble.  The fluid drag that the bubble encounters
is a function of the local CGM density $\rho(r)$ and the velocity of the blob
$\dot{r}$:
\begin{equation}
    \ddot{r} = -\rho(r)C_D\frac{\dot{r}^3}{|\dot{r}|}\frac{\pi R_{SB}(r)^2}{m_{SB}}
    \label{drag} 
\end{equation}
Where $C_D$ is the geometrically-dependent drag coefficient.  For a sphere, $C_D\sim
1/2$.  We note the dependence of this expression on the bubble radius $R_{SB}$, which thus needs to be solved for. If the CGM gas follows a power-law profile $\rho(r) =
\rho_0(r/r_0)^{-\beta}$, then we can derive a value for the bubble radius as it
rises, assuming it stays in pressure equilibrium with the surrounding medium:
\begin{equation}
    P = Kn^{5/3} \propto r^{\alpha-5\beta/3}
    \label{pressure}
\end{equation}
We assume that our CGM is in hydrostatic equilibrium, with a potential set by
the dark halo with negligible baryonic contribution (and assuming, as above, a
flat rotation curve with $v_c^2 = GM_{200}/R_{200}$, then we can use this to
determine $\beta$:
\begin{equation}
    \frac{dP}{dr} = -\nabla\phi\rho(r)
\end{equation}

\begin{equation}
    (\frac{5}{3}\beta-\alpha)K_{200}\rho_0^{2/3}r_0^{2\beta/3}r^{\alpha-2\beta/3} =
    \frac{GM_{200}}{R_{200}}
\end{equation}
Solving this equation yields the density profile slope $\beta = 3\alpha/2$.  We
can then integrate the density equation to determine the normalization of the
density profile.  If we assume that $f_{CGM}$ of the galactic baryons reside
within the CGM, then:
\begin{equation}
    \rho_0 = \frac{f_{B}f_{CGM}M_{200}}{4\pi R_{200}^3}(2-3\alpha/2)
\end{equation}

For the bubble to stay in pressure equilibrium with the CGM as it rises, it 
must grow from an initial radius of $R_{SB} \sim h$ to
\begin{equation}
    R_{SB}(r) = h^{1-3\alpha/10}r^{3\alpha/10}
    \label{SB_R_expanding}
\end{equation}

This finally allows us to determine the drag acceleration:
\begin{equation}
    \ddot{r} = \frac{-C_D f_B f_{CGM} M_{200}}{4m_{SB}}
    h^{1-3\alpha/10}r^{-6\alpha/5}R_{200}^{3\alpha/2-3}
    \frac{\dot{r}^3}{|\dot{r}|}
    \label{drag_full}
\end{equation}

Combining this with equation~\ref{acceleration_phys_nodrag}, we can finally
derive the full equation of motion that takes in to account the effects of
buoyancy, gravity, and fluid drag on this outflowing superbubble:
\begin{equation}
    \ddot{r} = \frac{v_c^2}{r}
    \left[\left(\frac{K_{SB}R_{200}^\alpha}{K_{200}r^\alpha}\right)^{3/5}-1\right]
    - \xi r^{-6\alpha/5}\frac{\dot{r}^3}{|\dot{r}|}
    \label{acceleration_phys_full}
\end{equation}
Where $\xi= -C_D f_B f_{CGM} M_{200} (4m_{SB})^{-1}
h^{1-3\alpha/10}R_{200}^{3\alpha/2-3} $ is the
constant-valued coefficient for drag set by the halo and the superbubble initial
conditions.  For the calculations that follow, we assume $f_{CGM}=0.5$, roughly
in agreement with the values of the CGM mass fraction observationally determined
by COS-Halos \citep{Werk2014}, and the amount required to explain the so-called
``missing baryons'' \citep{Gupta2012}.

The overall picture presented here is shown in schematic form in
figure~\ref{entropy_story}:  entropy generated by feedback lifts gas up through
the CGM, on a roughly adiabatic track.  Once that gas reaches its buoyant
equilibrium point, it may then cool radiatively, returning back to the ISM.
This cooling-feedback-buoyant uplift cycle will keep material in the CGM,
regulating the star formation, baryon masses, and metal content of galaxies.

\section{The trajectory of a buoyant bubble in a Milky-Way-Like CGM}
We can take the equations of motion from equation~\ref{acceleration_phys_full} and
integrate them forward in time numerically.  We use the scipy wrapper
\citep{scipy} for the ODEPACK LSODA method, which automatically switches between
an Adams-Bashforth and Backward Differentiation Formula (BDF) method when the
equations pass a stiffness criterion \citep{Hindmarsh1982}.  With this, we are
able to evolve the position and velocity of the buoyant wind over time.  

\begin{figure*}
    \includegraphics[width=\textwidth]{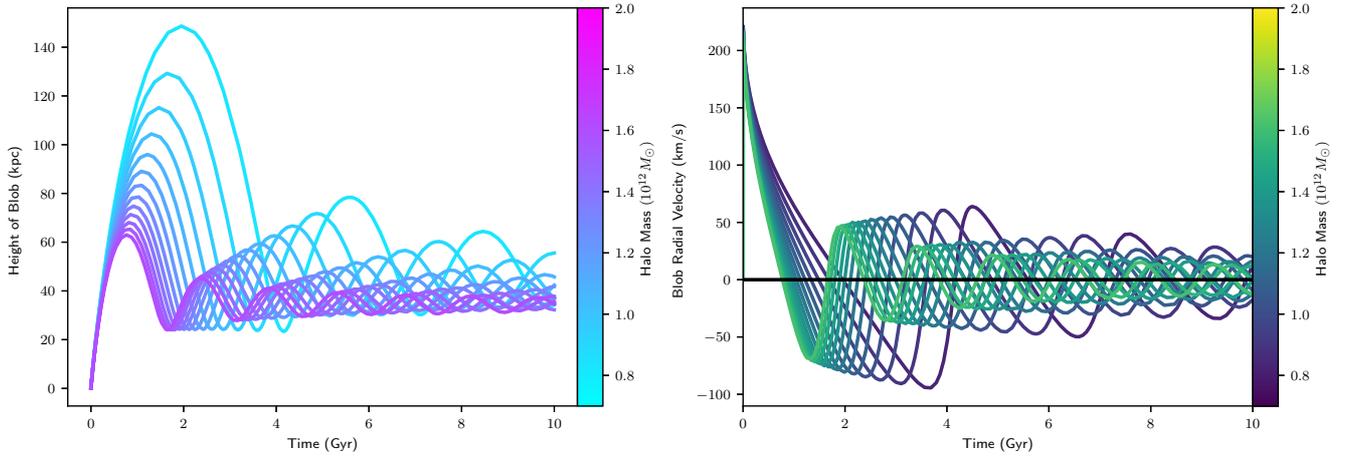}
    \caption{Flight of a buoyant bubble through the CGM. Left panel: Altitude in
    the galactic halo of a superbubble driven by a cluster with a mechanical
    luminosity of $10^{38}\;\rm{erg/s}$ ($M_{cluster} ~ 10^4\Msun$).  The
    cluster is embedded in a Milky Way-like ISM with a HI scale height of
    $200\pc$ and an averaged ISM density of $1\hcc$. The color of each curve
    denotes the virial mass of the galaxy.  Right panel: velocity of the same
    bubble. As can be seen, the bubble is rapidly accelerated during the
    first few $100\Myr$ of its flight through the CGM, and then subsequently
    loses velocity as it convectively overshoots and is damped through drag.} 
    \label{buoyant_flight}
\end{figure*}

We can start with a fairly simple case, to allow us to see the qualitative
behaviour of equation~\ref{acceleration_phys_full} integrated over time.  As
figure~\ref{buoyant_flight} shows, for a reasonably sized star cluster
($M\sim10^4\Msun$) in a Milky-Way like ISM with density $n_{amb}=1\hcc$ a scale
height of $200\pc$ \citep{Dehnen1998}, the apoapsis of the bubble's flight is
actually significantly higher than the buoyant equilibrium radius in the CGM
where it later oscillates about.  This phenomenon is convective overshooting,
occurring due to simple momentum conservation.  The bubble's oscillations
subsequent to this are damped fairly heavily by drag, as is shown in
figure~\ref{buoyant_flight}.  As that figure shows, these bubbles never exceed
velocities of $\sim200\kms$ in ~ $M*$ halos, despite being ejected to
$\sim100\kpc$ and re-accreting on timescales $>\Gyr$.  

\begin{figure*}
    \includegraphics[width=\textwidth]{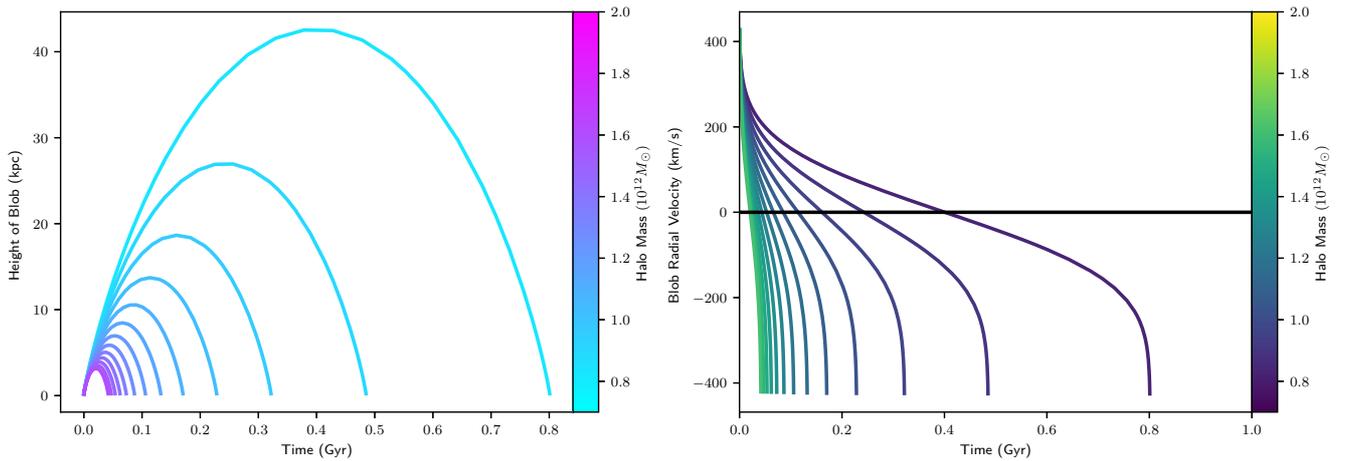}
    \caption{Flight of a ballistic bubble throught the CGM. Left panel shows
    galactocentric radius, right panel shows bubble velocity. Note that the time
    range on the horizontal axis is 10 times shorter than the previous figure. If, rather
    than allowing the hot interior of a superbubble to rise buoyantly, it is
    ejected ballistically, it will re-accrete on a much shorter timescale.  Here
    we see the trajectory of a parcel of gas assuming all of the star cluster
    feedback is deposited as kinetic energy into a mass set by
    equation~\ref{SB_mass}.  When buoyant uplift is unaccounted for, the
    fountain cycling time drops well below $\sim\Gyr$. When a bubble is launched
    ballistically, it begins with a fairly high ($>400\kms$) velocity, which it
    rapidly loses to gravity.  Note that here we {\it do not} include the
    effects of drag, effectively maximizing the ability for ballistically
    launched outflows to escape.}
    \label{ballistic_flight}
\end{figure*}

If we compare this behaviour to a kinetically launched, ballistic trajectory, we
see qualitatively that there is a stark difference between the two wind
launching mechanisms.  We calculate an initial velocity for the bubble assuming
that $100\%$ of the driving luminosity goes into kinetic energy, launching a
mass given by equation~\ref{SB_mass}.  In order to maximize the effects of this
ballistic wind, we do not include the drag terms given by
equation~\ref{drag_full}, allowing the bubble to be slowed by gravity alone.  We
see in figure~\ref{ballistic_flight} that unlike
entropy driven winds, these energy/momentum driven outflows initially travel
fairly fast ($\sim400\kms$) compared to the average in COS-Halos
\citet{Werk2014}. However, as this velocity is below the escape velocity for
the range of halo masses we examine, this material rapidly re-accretes on
timescales $<<t_{hubble}$, and only reaches modest heights above the disc.  If a
sufficiently smaller mass loading is assumed, this lower mass would of course
have a higher velocity. If that velocity exceeded the escape velocity of the
halo, would allow  the material to leave beyond the virial radius.  This is a
key difference between entropy driven winds and winds propelled ballistically:
entropy-driven winds do not ``fail'' in the same way as ballistically driven
winds. Even with their lower characteristic velocities, entropy-driven winds
persist in the CGM long after ballistically propelled outflows would have
re-accreted. 

If we consider an arbitrary mass-loading $\eta$, then the initial velocity of
this wind, assuming $100\%$ of the energy deposited by supernovae is coupled to
the kinetic energy of this outflowing material, we can see that
$v_0\sim700\eta^{-1/2}\kms$.  For an NFW halo the escape velocity is $v_{esc} =
\sqrt{6c}v_{200}\sim800\kms(M_{200}/10^{12}\Msun)^{1/3}$ (assuming a
concentration of $c\sim4$, \citealt{Zhao2003}). This means that for outflow mass
loadings to be significant $\eta>1$, and not re-accrete on short $\sim100\Myr$
timescales, these outflows cannot be driven ballistically.  Entropy-driven
winds, on the other hand, naturally give high mass loadings with long recycling
times.

\begin{figure}
    \includegraphics[width=0.5\textwidth]{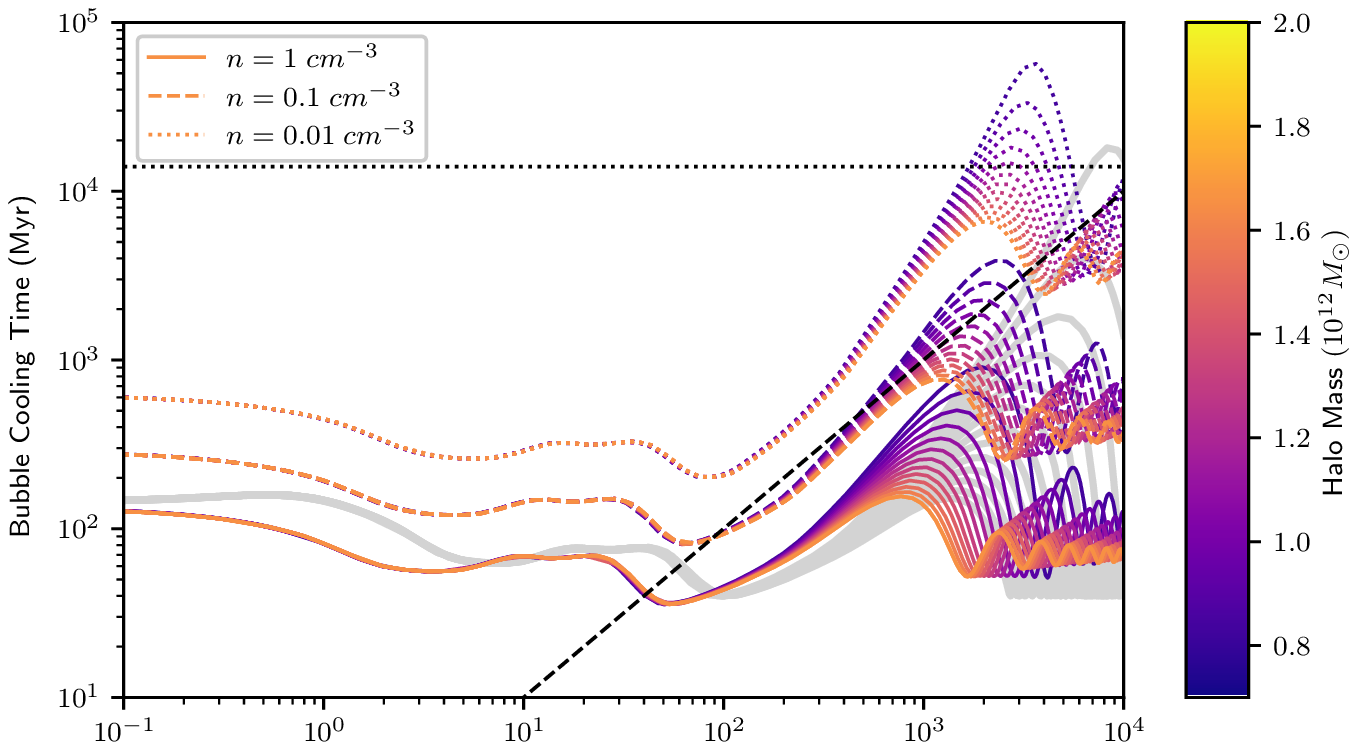}
    \includegraphics[width=0.5\textwidth]{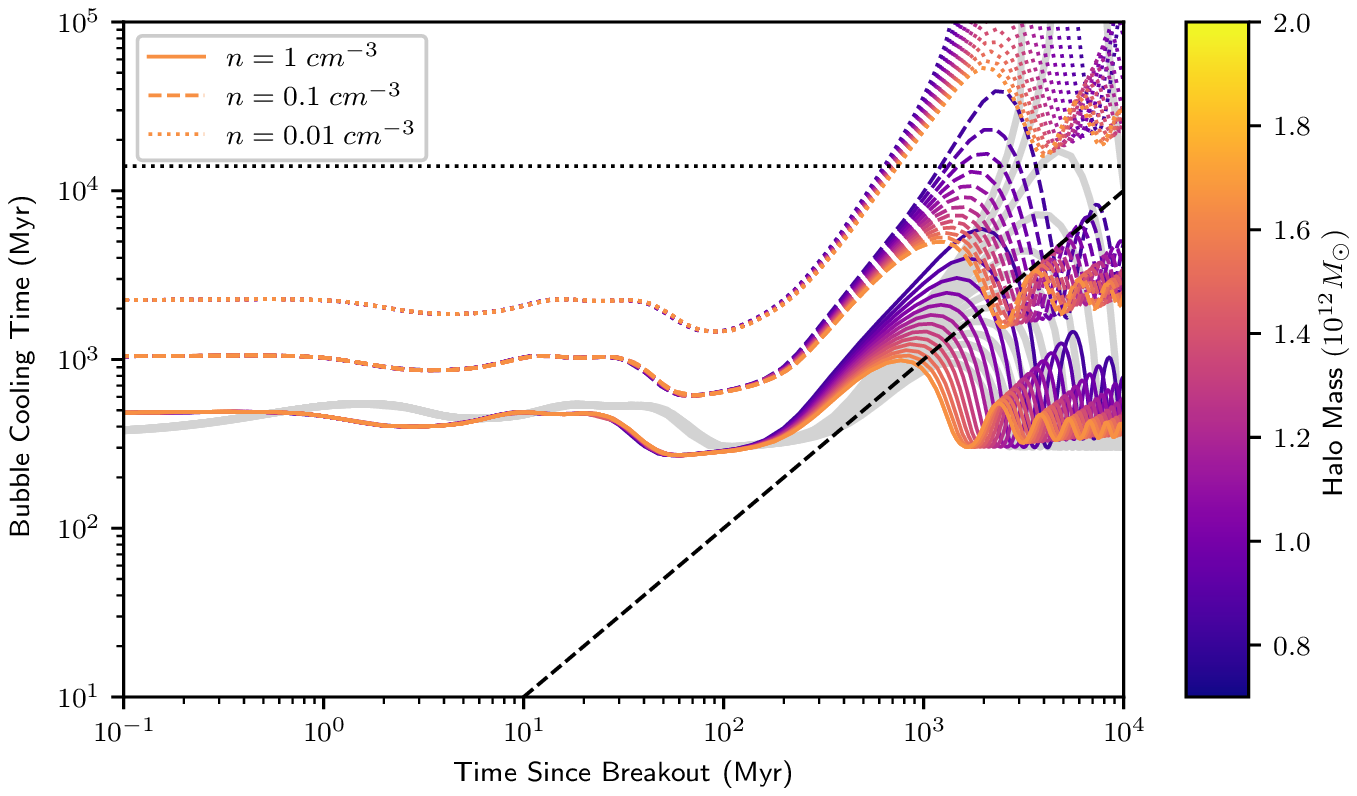}
    \caption{Cooling time of a hot bubble in as it rises through the CGM.  The
    top panel shows the cooling times for a metallicity of $Z_\odot$, while the
    bottom shows cooling times for a metallicity of $0.1Z_\odot$. The three
    different line styles (solid, dashed, and dotted) show cooling times for
    superbubbles driven in an ISM with ambient density of $1\hcc$,
    $10^{-1}\hcc$, and $10^{-2}\hcc$ respectively.  The horizontal dotted line
    shows a Hubble time, and the dashed black curve shows
    the $1:1$ line, where the cooling time is equal to the bubble's age, and
    shows roughly when the bubble will lose its entropy to radiative cooling.
    The light grey curves show how, paradoxically, increasing the SN luminosity
    of the cluster by a factor of 10 doesn't significantly change the time when
    it cools.  As this figure shows, pre-processing of the ISM by early feedback
    or past SN explosions, thus lowering the ambient density superbubbles
    detonate in, is essential to producing superbubbles with sufficiently long
    cooling times to rise through the CGM without radiating away their entropy.}
    \label{cooltime_Mvir}
\end{figure}
\begin{figure*}
    \includegraphics[width=\textwidth]{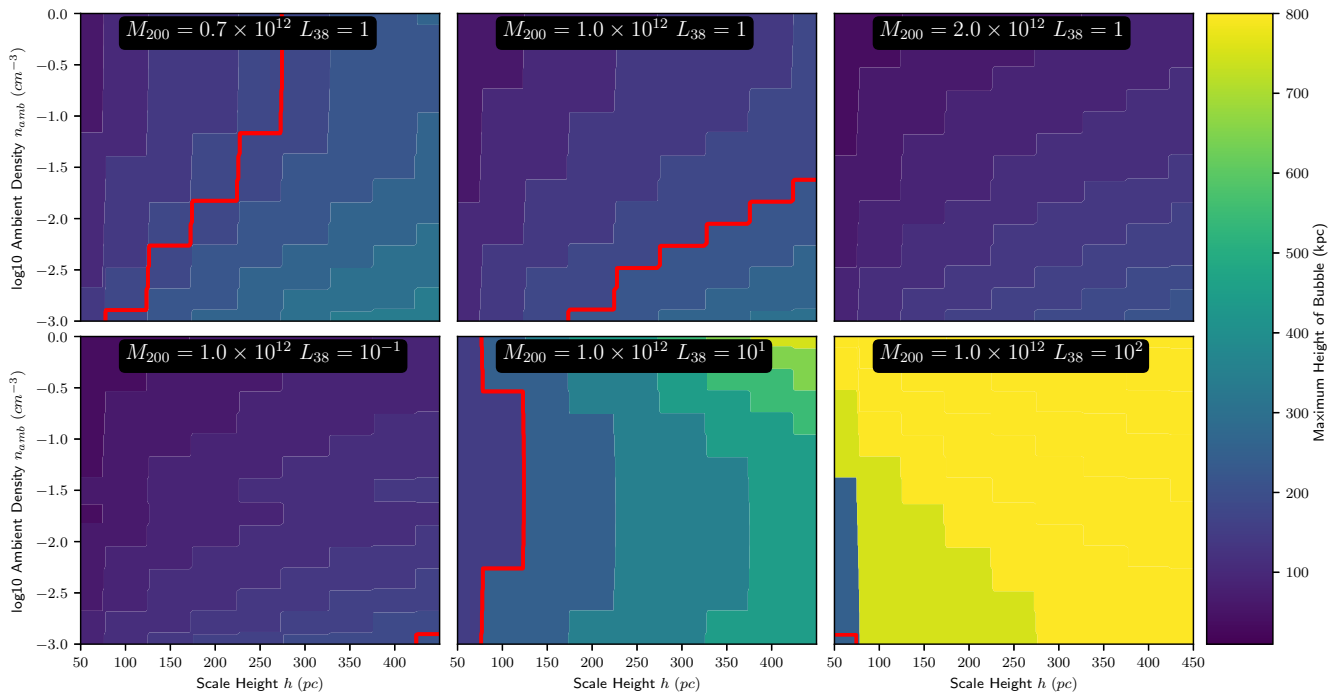}
    \caption{Contour plot of the maximum height (within the $10\Gyr$ of
    integration time we use) obtained by a buoyant bubble launched from a number
    of different ISM conditions, by a range of driving luminosities, for 3
    different halo masses.  The top row shows bubbles driven by a cluster with
    luminosity $10^{38}\ergs$, for halo masses of $7\times10^{11}\Msun$,
    $10^{12}\Msun$, and $2\times10^{12}$ (left to right).  The bottom row shows
    the result of a cluster with $10^{37}\ergs$, $10^{39}\ergs$, and
    $10^{40}\ergs$ (left to right) in a halo with $M_{200}=10^{12}\Msun$.  The
    red contour shows the virial radius of the halo.  As would be expected from
    equation~\ref{SB_entropy}, bubbles escaping from thicker, lower-density ISM
    reach higher through the CGM, and bubbles rising through the CGM of more
    massive haloes are trapped to lower heights.  Even bubbles which do not
    reach the virial radius of their haloes still can rise $\sim100\kpc$ in all
    but the most massive haloes, with the densest, thinnest ISM.  The black
    lines show when the bubbles have at $10\Gyr$ still not yet reached their
    highest altitude.}
    \label{apsis}
\end{figure*}
\begin{figure*}
    \includegraphics[width=\textwidth]{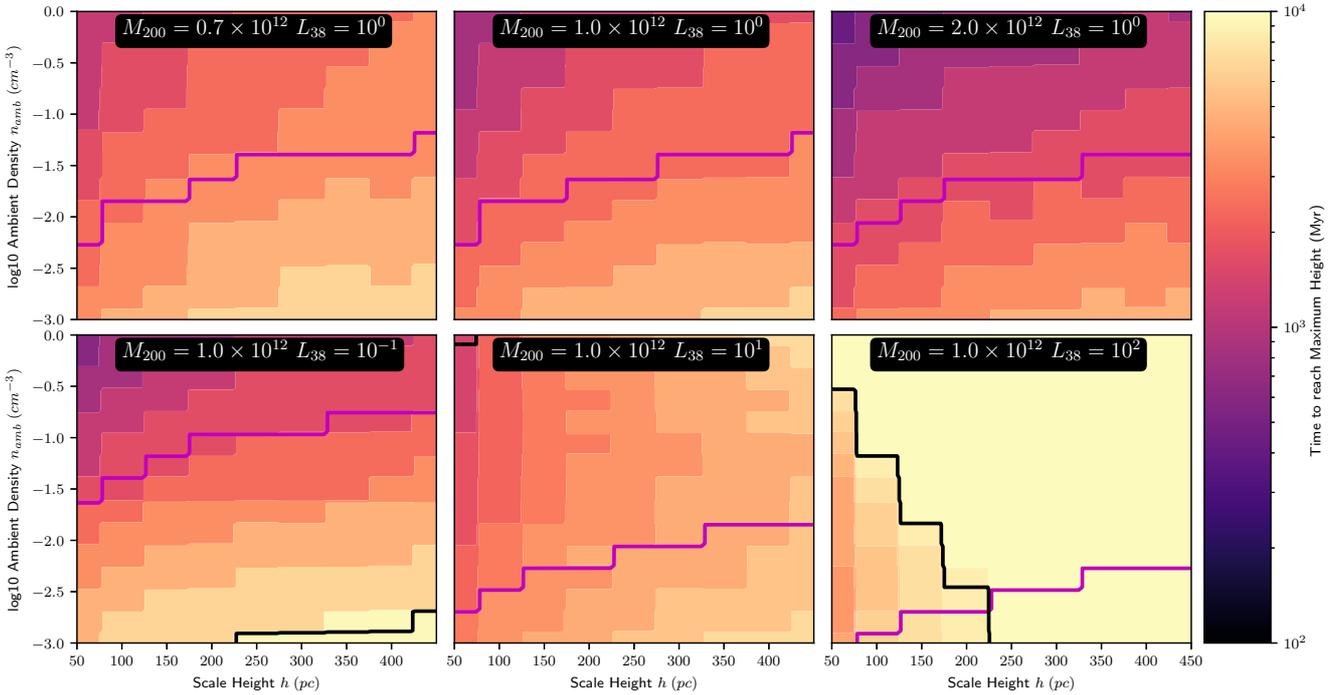}
    \caption{Contour plot of the time required to reach maximum height, for a
    bubble launched from a number of different ISM conditions, for the same
    parameters as figure~\ref{apsis}.  As can be seen here, even for the bubbles
    which reach the lowest heights, they can rise through the CGM for a
    $\sim\Gyr$  The purple line shows the time when
    the bubble will have lost its entropy through radiative cooling (the region
    above this line cools before it reaches its maximum height).  As this
    shows, the primary determinant of whether a bubble will reach the peak of
    its buoyant trajectory before cooling is whether or not it is launched from
    a sufficiently low density ambient medium.  The black curve shows where the
    bubble has not yet reached the maximum height of it's flight by the end of
    our $10\Gyr$ integration.}
    \label{t_apsis}
\end{figure*}
\begin{figure*}
    \includegraphics[width=\textwidth]{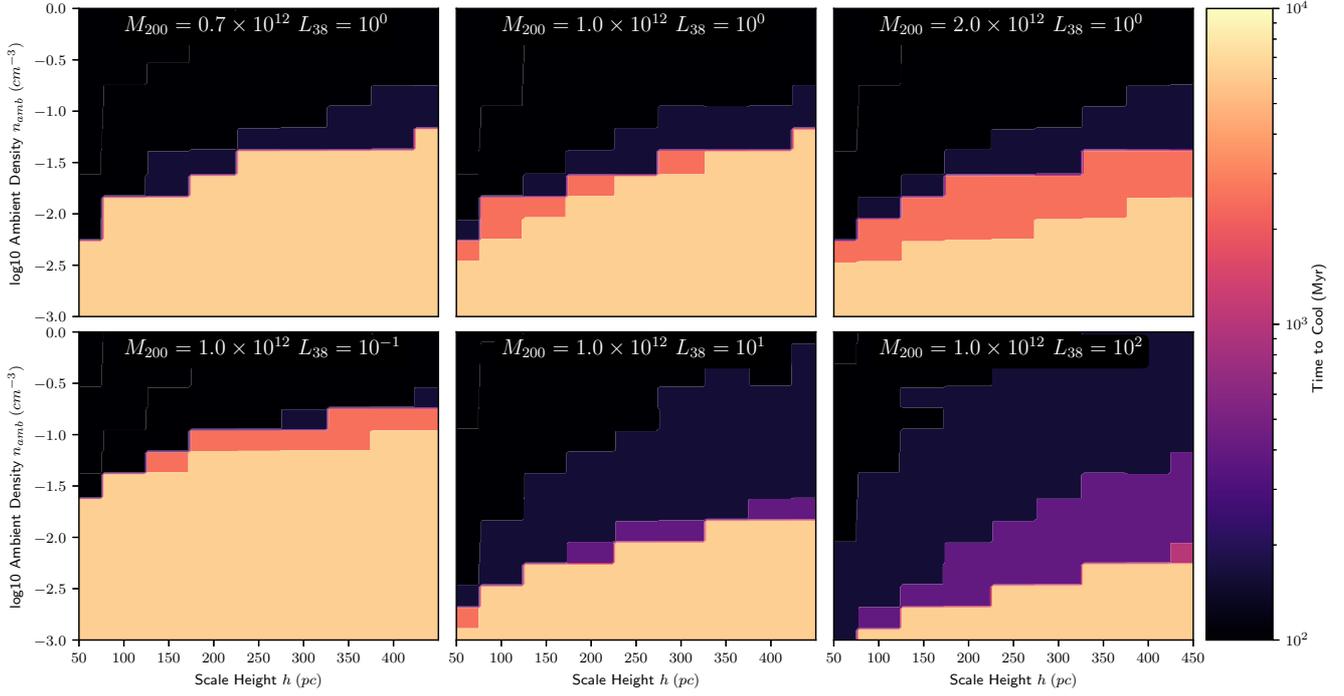}
    \caption{Contour plot of the time when a bubble will radiatively cool, for
    the same parameters as figure~\ref{apsis}.  As was seen in
    figure~\ref{t_apsis}, the density of the ambient medium is the most
    important factor in determining when the bubble will cool, and for different
    halo masses/driving luminosities, this density ranges between
    $\sim10^{-1}\hcc$ and $\sim10^{-3}\hcc$.  Paradoxically, bubbles driven by
    higher luminosity clusters actually require {\it lower} ambient ISM
    densities to avoid cooling, because while they result in hotter bubbles,
    they also result in denser ones at breakout, as $n_{SB}\propto L^{8/21}$.}
    \label{t_cool}
\end{figure*}
\begin{figure*}
    \includegraphics[width=\textwidth]{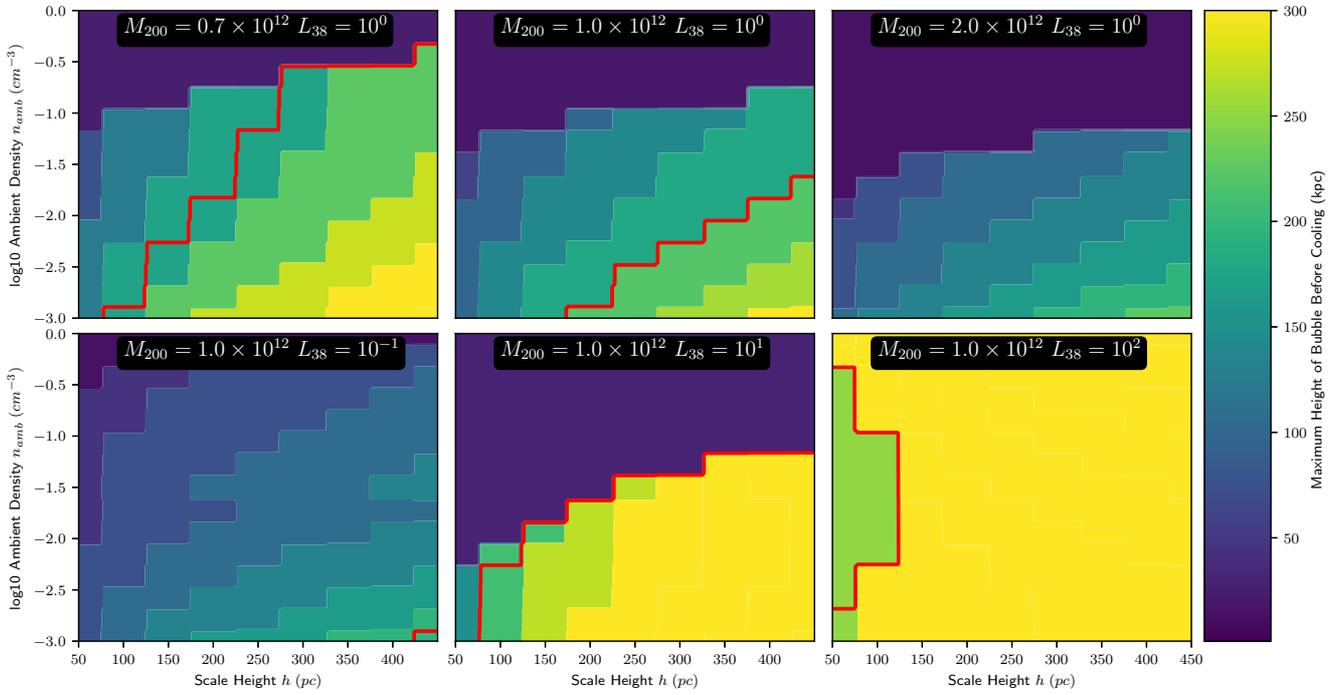}
    \caption{If we assume the bubble's true maximum height is the maximum height
    it obtains prior to cooling, we can see that, except in the case of our most
    massive haloes, bubbles are still able to reach appreciable $\sim100\kpc$
    heights before they radiatively cool.}
    \label{apsis_cool}
\end{figure*}
\begin{figure*}
    \includegraphics[width=\textwidth]{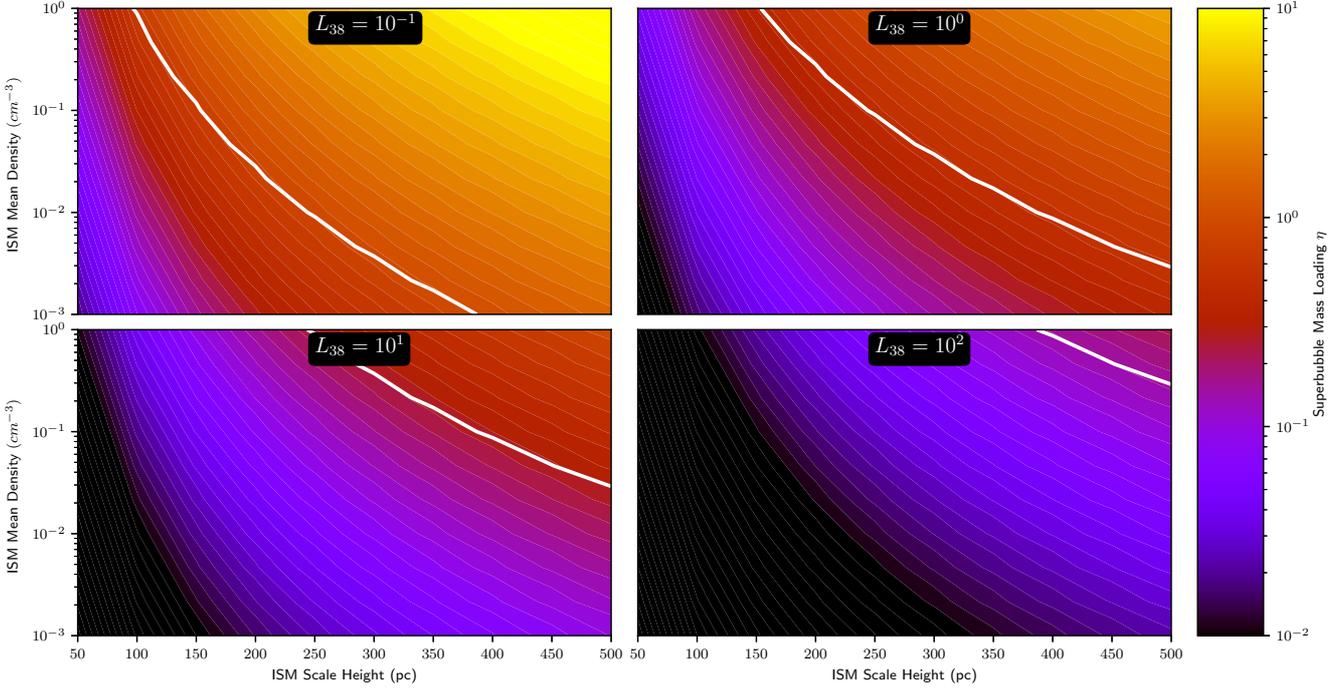}
    \caption{Mass loading ($\eta=M_{SB}/M_{cluster}$) for different mass
    clusters ($10^3\Msun$ in the upper left, $10^4\Msun$ in the upper right,
    $10^5\Msun$ in lower left, and $10^6\Msun$ in lower right).  The white curve
    shows the region where a bubble will breakout before the cluster shuts off
    it's feedback (as in figure~\ref{SB_breakout_times}).  Bubbles driven in
    regions to the right of this curve will not break out before the end of SN
    feedback.  As can be seen here, for most clusters in a reasonable ISM, the
    mass loadings are modest, with $\eta<1$.  Larger mass loadings will be
    driven only from a dense, thick ISM.}
    \label{SB_massloading}
\end{figure*}
\begin{figure}
    \includegraphics[width=0.5\textwidth]{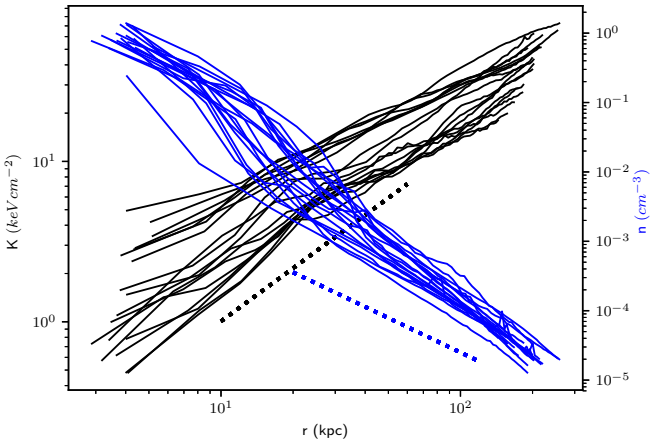}
    \caption{Entropy  and density profiles for the 18 MUGS2 galaxies.  The left
    panel shows the volume-averaged entropy for each galaxy in black, while the
    right panel shows the volume-averaged density profile for each galaxy in
    blue.  The dotted curves show the fiducial $\alpha=1.1$ slope for the
    entropy profile, and the corresponding hydrostatic $\beta=3/2\alpha$ slope
    for the density profile.  While the inner part of the density profile is
    noticeably steeper than the hydrostatic solution (in large part due to the
    contribution of the disc), the outer parts of the halo do appear to be in
    rough hydrostatic equilibrium with a power-law entropy slope of
    $\alpha=1.1$.}
    \label{halo_entropy}
\end{figure}
\begin{figure}
    \includegraphics[width=0.5\textwidth]{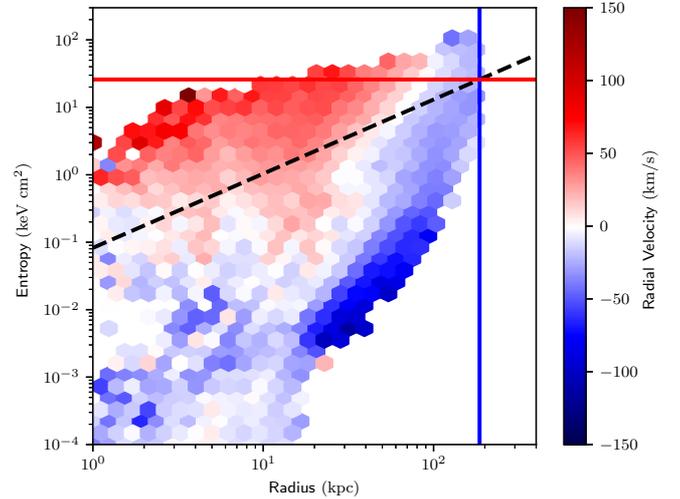}
    \caption{Mean radial velocities as a function of entropy and radius.  As in
    figure~\ref{entropy_story}, we show the virial entropy with the horizontal
    red line and the virial radius with the vertical blue line.  The CGM
    entropy profile, as given by equation~\ref{CGM_entropy}, is shown with the
    black dashed line.  As can be seen here, the outflow velocity generally
    increases as the entropy exceeds the CGM entropy profile, while the inflow
    velocity increases the further below the entropy profile a parcel resides
    (and the closer it gets to the bottom of the potential well).}
    \label{entropy_velocities}
\end{figure}
\begin{figure}
    \includegraphics[width=0.5\textwidth]{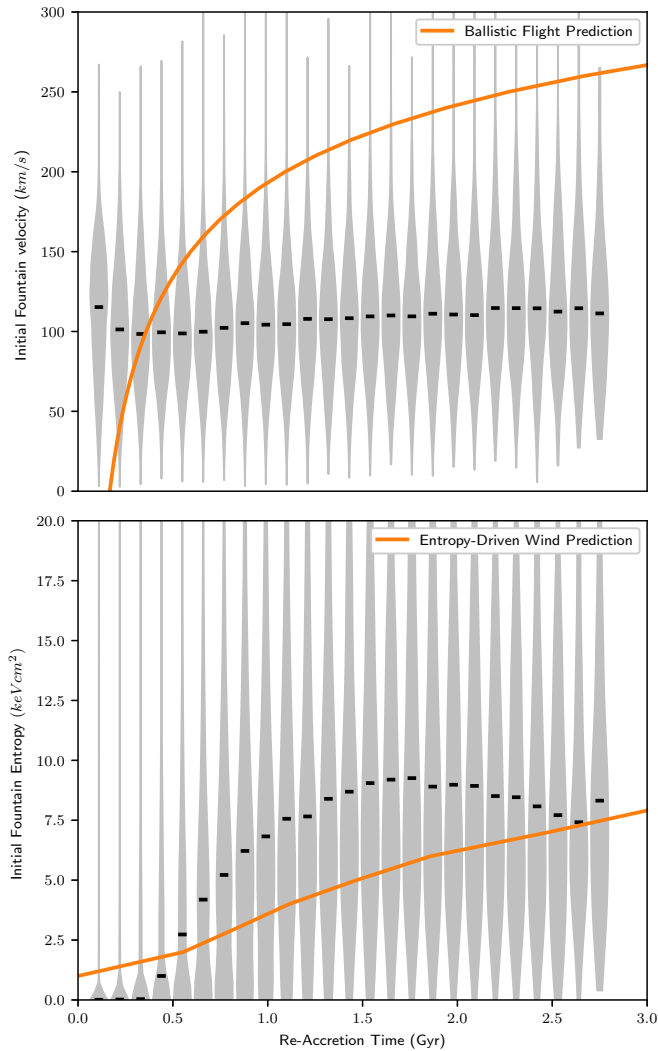}
    \caption{In a simulated $L*$ galaxy near $z=0$, outflow re-accretion times
    for fountain gas have almost no relation to the velocity at ejection, with
    most fountains having an initial velocity of $\sim100\kms$.  If we instead
    look at the typical entropy of ejected gas, we can see that longer-lived
    $>2\Gyr$ fountains are dominated by gas with higher entropy
    $\sim5-10\mathrm{kev\;cm^2}$.  Median values for the fountain initial velocity
    and entropy are shown with black lines, and the predictions for the flight
    of a ballistic and entropy-driven fountain are shown with the orange
    curves, assuming a gas scale height of $200\pc$ and a mean ISM density of
    $1\hcc$, driven by a cluster with a mechanical luminosity of $10^{38}\ergs$}
    \label{fountain_time}
\end{figure}
\begin{figure*}
    \includegraphics[width=\textwidth]{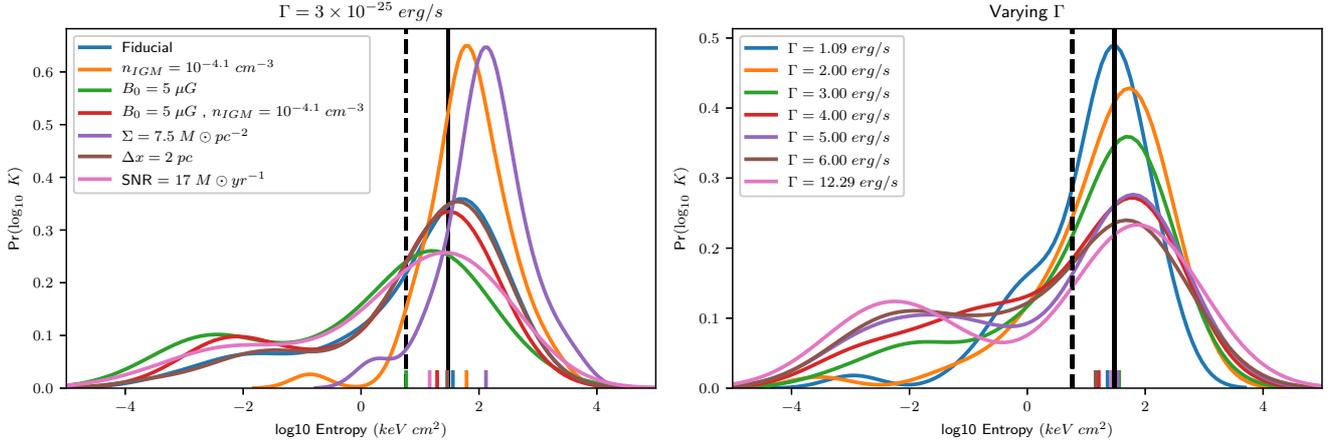}
    \caption{Entropy distribution kernel density estimation of the inner $20\pc$
    of superbubbles in 14 different simulations from \citet{Hill2018}.  The left
    hand panel shows variations in the resolution, supernova rate, ISM and IGM
    density, and magnetic field properties, while the right hand panel shows the
    results of varying the intensity of UV heating on the ISM, with the same
    properties as the fiducial simulations in the left panel.  As can be seen,
    every case produces a significant amount of high-entropy gas, even above the
    $30\;{\rm keV\;cm}^2$ (shown as the solid black vertical line) required to
    be neutrally buoyant at the virial radius of a $10^{12}\Msun$ halo, and well
    above the $5.8\;{\rm keV\;cm}^2$ normalization in equation~\ref{SB_entropy}
    (shown as the vertical dashed black line).  The right hand panel shows that
    higher heating rates $\Gamma$ result in a broader distribution of entropies,
    likely due to a larger fraction of the ISM existing in intermediate
    densities between the cold and hot phases.  As might be expected, both
    magnetic fields and lower SN rates reduce median entropy produced by
    superbubbles, while a lower gas surface density or a cooler, denser CGM
    results in much higher entropy bubbles.  The rug plot at the bottom of the x
    axis shows the median entropy for each simulation.  As this shows, there is
    much greater variation in the median entropy in the left hand panel.}
    \label{Hill_entropy}
\end{figure*}
\begin{figure}
    \includegraphics[width=0.5\textwidth]{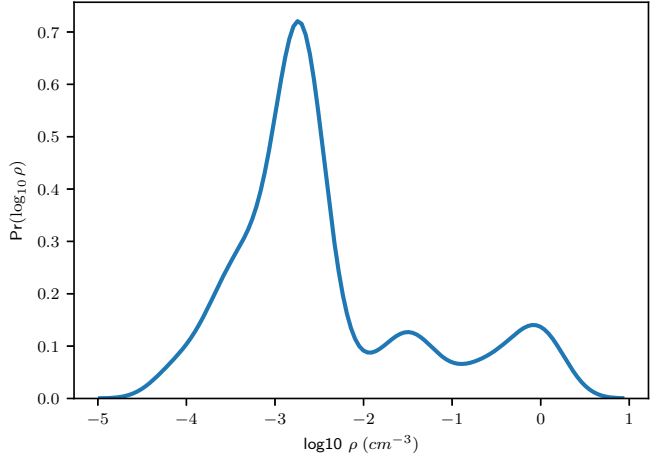}
    \caption{Density distribution of the $20\pc$ sphere surrounding each
    pre-supernovae star particle in the \citet{Hill2018} simulations.  As is
    clear here, the majority of SN detonate in a region with significantly
    lower density than $1\hcc$, with most occurring in regions with densities
    $\sim10^{-3}\hcc$.  This is due to a combination of stellar winds
    pre-treating the ISM, heating and disrupting dense gas; and the random
    positioning of star clusters, so that many detonate in the low density,
    hot superbubbles of older star clusters.}
    \label{Hill_density}
\end{figure}
\begin{figure*}
    \includegraphics[width=\textwidth]{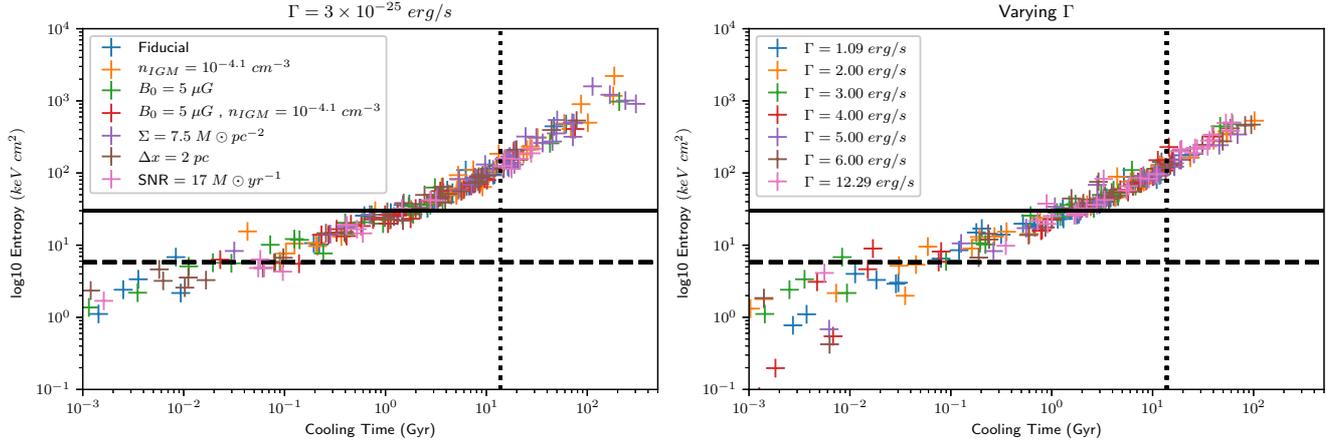}
    \caption{Cooling times versus entropies for each superbubble in the
    \citet{Hill2018} simulations.  The vertical dotted line shows a Hubble time,
    and the two horizontal lines show the viral entropy of $10^{12}\Msun$ halos
    $30\;{\rm keV\;cm}^2$ (solid line) and the typical bubble entropy at
    breakout, $5.8\;{\rm keV\;cm}^2$ (dashed line) respectively.  As is clear,
    essentially every superbubble with an entropy above $5.8\;{\rm keV\;cm}^2$
    has a cooling time longer than $100\Myr$, and the majority of superbubbles
    have cooling times exceeding $1\Gyr$, as would be expected from
    figure~\ref{cooltime_Mvir} and figure~\ref{Hill_density}.  Superbubbles with
    short cooling times would already not be able to rise buoyantly (due to
    their low entropy) even if they could break out of the ISM before cooling.}
    \label{Hill_cooltime}
\end{figure*}
\begin{figure}
    \includegraphics[width=0.5\textwidth]{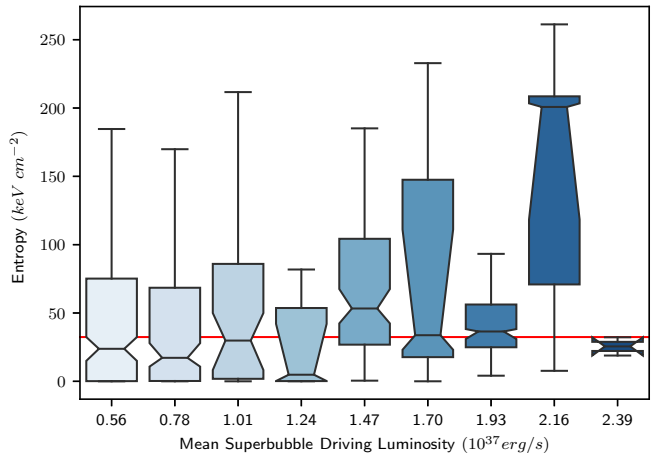}
    \caption{Box plot of the entropy distribution of superbubbles binned as a
    function of the SN luminosity of their star clusters.  The red horizontal
    line shows the median entropy of $32\;{\rm keV\;cm}^2$.  The box heights are
    the interquartile range, while the whiskers show 1.5 times the IQR.  The
    notches show the bootstrapped $95\%$ confidence interval for the median
    values in each bin. No clear trend with luminosity is obvious for this range
    of cluster masses.}
    \label{Hill_clustering}
\end{figure}
\begin{figure}
    \includegraphics[width=0.5\textwidth]{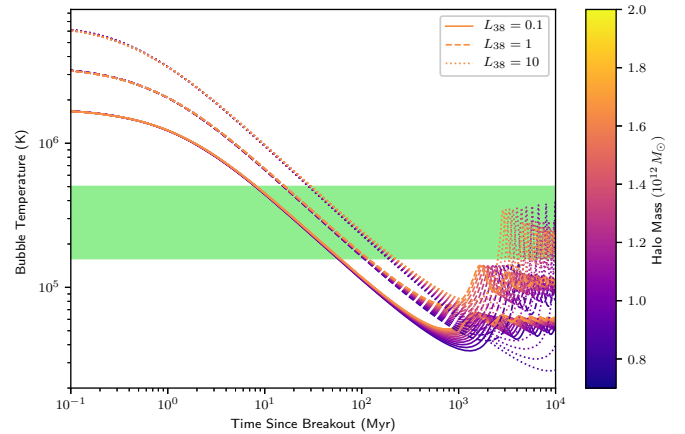}
    \caption{Temperature of the bubble as it rises through the CGM.  The green
    band shows the region where the {\sc Ovi} ionization fraction exceeds
    $f_{OVI}=0.02$ \citep{Tumlinson2011}. While the bubble spends most of it's
    lifespan below the critical temperature $T^{5.2-5.7}\K$ where {\sc Ovi} is
    abundant, it does spend a few $10-100\Myr$ during it's flight in this
    temperature range, where it in principle should be detectable in absorption by COS and
    other UV instruments.  For bubbles driven by more massive clusters, the
    oscillations about buoyant equlibrium may also push the bubble temperature
    back towards this temperature as it heats and cools through PdV work.}
    \label{temp_Mvir}
\end{figure}
\begin{figure}
    \includegraphics[width=0.5\textwidth]{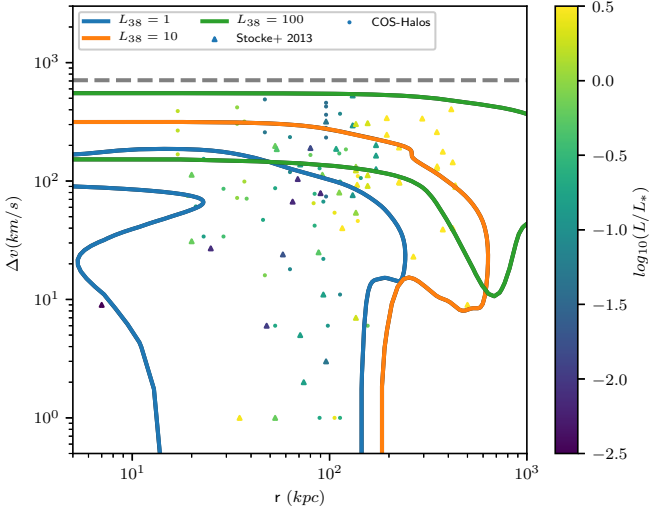}
    \caption{Here we show a comparison between the velocities and positions of
    entropy-driven wind bubbles and observations from the HST-COS instrument.
    We show contours outlining the regions in phase space where the bubble finds
    itself for 3 different driving luminosities in a $7\times10^{11}\Msun$ halo.
    As can be seen here, a range of cluster masses can explain all of the
    structure we see, and in fact, different ranges of this phase space may
    point to driving by different luminosity sources.  Data points are from
    \citet{Stocke2013} and \citet{Tumlinson2013}, shown with triangles and dots
    respectively, coloured by their host galaxy luminosity.  The grey dashed
    line shows the escape velocity for a halo with virial mass
    $7\times10^{11}\Msun$, $\sim700\kms$.}
    \label{velocity_radius}
\end{figure}
\begin{figure}
    \includegraphics[width=0.5\textwidth]{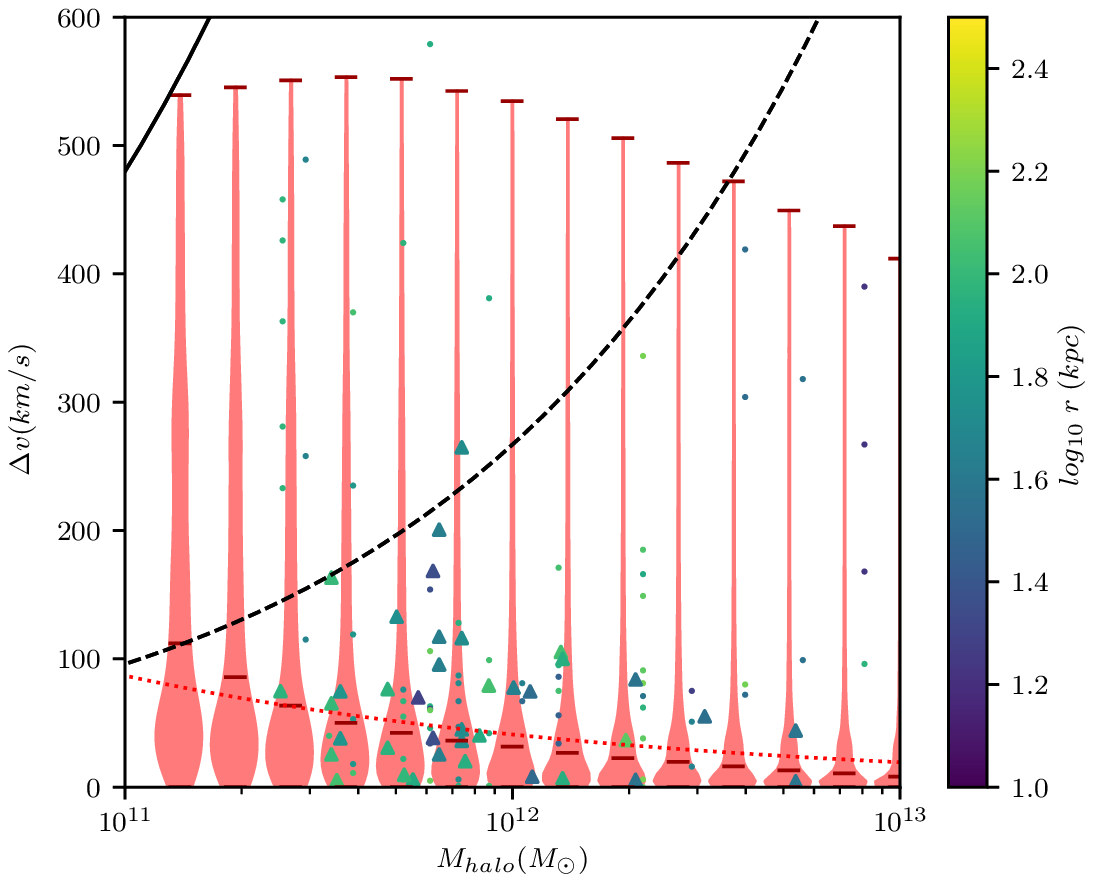}
    \caption{Outflow velocity as a function of  halo mass.  Data points are from
    COS-Halos, either the full data set (shown as small circles)
    \citep{Tumlinson2013}, or from the subsample of highly ionized sources from
    \citep{Werk2016} (large triangles).  Predictions for the velocity of
    entropy-driven winds from a range of ISM densities ($10^{-3}-1\hcc$) and
    scale heights ($50-500\pc$) for different halo masses are shown as violin
    plots.  The horizontal lines in the violin plot show the most extreme data
    points (top) and median values (bottom).  The dotted red curve shows a power-law fit to the
    highly ionized COS-Halos sources.  The solid and dashed black curves show
    the escape and virial velocities respectively.}
    \label{cos_haloes}
\end{figure}

\subsection{Cooling of the Hot Bubble}
Radiative cooling of the hot interior of a superbubble will remove entropy from
that gas. If the cooling time of the superbubble is short compared to the time
it spends cycling through the CGM, it will lose buoyancy and fall back to the
ISM.  However, as it rises through a progressively lower-pressure CGM it will
expand adiabatically, decreasing both its density and temperature.  This has two
opposing effects: decreasing the cooling rate due to lower density, while
increasing the cooling rate as it approaches the recombination peak of the
cooling curve at $10^5\K$ \citep{Raymond1976}.  Simply calculating the cooling
time of the bubble at break out is insufficient to determine whether it will
evolve adiabatically as it rises through the CGM.

We therefore instead use the trajectories of bubbles rising through the CGM,
calculated by numerically integrating equation~\ref{acceleration_phys} to obtain
temperatures and densities for superbubbles launched from a number of potential
ISM environments.  From equation~\ref{SB_Tmean} and
equation~\ref{SB_breakout_time}, we can see that the temperature of the bubble
at break out is independent of the ambient density, as $n(0) \propto
n_{amb}^{1/3}$ and $t_7 \propto n_{amb}^{1/3}$, while $T \propto t_7/n(0)$.
This temperature should scale weakly with scale height, $T \propto h^{-3/7}$.
The density of the bubble will scale sub-linearly with ambient density, with
$n_{SB} \propto n_{amb}^{1/3}$ and  will be nearly linear with the inverse of
the ISM scale height $n_{SB} \propto h^{-22/21}$.  Since the primary cooling
processes are collisional at the temperatures of this bubble, and therefore
$\Lambda \propto n_{SB}^2$, this suggests that the ambient density $n_{amb}$ is
the primary parameter for determining whether the superbubble will cool, unless
we approach the peak of the cooling curve, where $|\frac{d\Lambda}{dT}|$ becomes
large.

In order to investigate the evolution of the cooling times with more rigour, we
used the {\sc Grackle} cooling library \citep{grackle} to calculate the cooling
times of the bubble at each timestep as its trajectory through the CGM is
integrated.  These cooling times can be compared to the timescale required for
the bubble to travel significant distances from the disc, and as an estimate for
the timescale over which the bubble can continue to oscillate in the CGM.  We
assume that each bubble has solar metallicity $Z=0.0129$, and that it is subject
to a \citet{Haardt2012} UV background field.  We vary the background density and
ISM scale height, as well as the SN luminosity of the driving cluster.  As
figure~\ref{cooltime_Mvir} shows, while the initial cooling time of the bubble
scales as we expect, sublinear with the ambient ISM density, the subsequent
evolution, especially as the bubble reaches it's apsis, is sensitive to both the
ISM scale height and the ambient density.  For metallicities below solar,
however, the cooling rates are supressed by the lack of metal emission lines,
and with a reduction to just $Z=0.1Z_\odot$, even clusters in a dense ISM will
be able to drive bubbles that take $\sim\Gyr$ to cool.  This means that
entropy-driven winds would have been more effective at higher redshift, when the
cosmic metallicity was lower, and even more effective than the scalings relative
to halo mass we will see with constant (solar) metallicity, as the
mass-metallicity relation \citep{Erb2006,Ma2016} implies lower mass galaxies
will have lower metallicities.  These galaxies will thus not only have lower
virial entropies, but superbubbles driven from their ISM will have longer
cooling times.

\subsection{How Far Can Buoyant Outflows Travel?}
It is clear from the previous sections (and of course, from the derived
equations) that the evolution of buoyant bubbles through the CGM has dependence
on a number of factors: the radiative cooling of gas, the ISM scale height and
density, the driving cluster mass/luminosity, and the mass of the halo.  In this
section, we explore the sensitivity of outflows to these parameters.  We
integrate our equation of motion, equation~\ref{acceleration_phys} over a
4-dimensional grid spanning $L_{38}=0.1-100$, $h=50-500\pc$,
$n_{amb}=10^{-3}-1\hcc$, and $M_{200}=0.7-2\times10^{12}\Msun$.  We also use
{\sc Grackle} to calculate cooling times for each point in our integration.

Figure~\ref{apsis} shows the maximum apoapsis (height) of the bubble during an
integration time of $10\Gyr$.  This height is always the first turnover point in
the flight (as subsequent oscillations are damped by drag), except in the case
where the apoapsis time is $>10\Gyr$.  A number of features are clear from these
panels.  First, as the virial mass increases, the maximum height obtained by the
bubble is, unsurprisingly, decreased.  For our lightest halo, most of the
parameter space allows for a $10^4\Msun$ cluster to launch an entropy-driven
outflow that will escape beyond the virial radius.  Secondly, the ISM that
maximizes the bubble's apoapsis is one which is thick and diffuse, with a low
ambient density and large scale height.  Finally, as the mass of the cluster
increases, the maximum distance the subsequent outflow can reach increases as
well.

The time that these outflows reside in the CGM, more so than the distance they
reach, is critical to actually regulating the star formation, black hole
growth, and baryon content of the galactic disc.  Only by spending a
significant fraction of the galaxy's lifetime in the CGM can entropy-driven
outflows suppress star formation and bulge/black hole growth.  In
figure~\ref{t_apsis}, we show the time that a bubble takes to reach the maximum
height (this time can be thought of as $\sim1/2$ of the re-accretion time, if
we assume that the first periapsis drives the material back onto the ISM).  As
this figure shows, for essentially all of the configuration space we have
examined, entropy-driven winds can produce outflows that take $>\Gyr$ to
re-accrete.  This means that a parcel of fluid need only be ejected a handful
of times to spend most of its life outside of the star forming disc, as opposed
to the hundreds of times required for a mass-loaded, kinetically driven wind in
a $2\times10^{12}\Msun$ halo.  Even when radiative cooling is considered, as
long as the ambient ISM is sufficiently diffuse, either due to gas exhaustion
of the natal molecular cloud \citep{Kruijssen2012a,Longmore2014,Ginsburg2016} or
pre-treatment by early stellar feedback \citep{Chevance2020}, the bubble will
still spend a significant fraction of $t_{hubble}$ in the CGM.  We show the
radiative cooling times in figure~\ref{t_cool}.  As this shows, there is a
sharp distinction between bubbles with short $(\sim100\Myr)$ and long $(>\Gyr)$
cooling times.  This dividing region is sharpest in low-mass haloes, where the
bubble reaches greater heights and experiences the largest amount of adiabatic
cooling.  The most important factor is not the halo mass, however, but the
ambient density, and to a lesser extent, the ISM scale height.  These two
factors determine both the temperature and density of the bubble at break out.
A more diffuse ambient medium gives a cooler, more diffuse bubble, as does a
larger ISM scale height.  With a diffuse enough ambient ISM, the cooling time
of the bubble will be sufficient for it to both reach large apocentric
distances and stay out of the star forming-ISM.

Indeed, when we look instead at the maximum height a bubble obtains while it is
still evolving adiabatically, in figure~\ref{apsis_cool}, we can see that for
our most massive halo $2\times10^{12}\Msun$, only very large clusters, driving
superbubbles in extremely diffuse, thick discs can actually drive outflows to an
appreciable distance through the CGM, reaching just a few times $10\kpc$ before
radiative losses sap them of their entropy.  For a $< 10^{12}\Msun$ halo, the
largest star clusters can easily eject material beyond the virial radius.
The somewhat stronger dependence on cluster mass ($L_{38}$) that we see here
compared to figure~\ref{apsis} is due to the additional influence of radiative
cooling.  Without radiative losses, the relative importance of $L_{38}$ is
weaker (as we would expect from equation~\ref{SB_entropy}).  When radiative
losses are included, the higher initial SB temperature allows the bubble to
retain its entropy for longer, reaching higher altitudes in the CGM.
Though the lowest mass clusters only loft material to $\sim100\kpc$,
figure~\ref{t_cool} shows this material will stay held in the CGM by buoyancy
for $>\Gyr$ if the SB detonate in regions that have density below
$n_{amb}\sim0.1\hcc$.  As we will will see later, this is not a particularly
stringent criterion.

\subsection{Mass Loading and Mixing}
Whether entropy-driven winds can actually remove a significant amount of mass
from the galactic ISM depends critically on how much material is actually
carried by the bubble.  In the limit where the only mixing that occurs between
the ISM and the hot bubble is the evaporation captured in
equation~\ref{SB_density}, the mass loading value $\eta$ is simply
$\eta=m_{SB}/M_{cluster}$.  We can use equation~\ref{SB_mass}, along with
$L_{38}\sim M_{cluster}/10^4M_\odot$, to calculate this:
\begin{equation}
    \eta = 1.18 \left(\frac{n_{amb}}{\hcc}\right)^{1/3}
    \left(\frac{M_{cluster}}{10^4\Msun}\right)^{-13/21}
    \left(\frac{h}{267\pc}\right)^{41/21}
    \label{SB_eta}
\end{equation}
This shows that, a more massive cluster will give lower mass loadings than a
less massive one.  This is simply because a more massive cluster will produce a
bubble with density that scales sub-linearly with the cluster mass
($M_{SB}\propto M_{cluster}^{8/21}$).  As we show in
figure~\ref{SB_massloading}, for most of the range of ISM densities and scale
heights we consider, the mass loading given by equation~\ref{SB_eta} is between
$\eta\sim0.01-10$.  Only relatively low-mass clusters driving bubbles in a
dense, thick ISM can produce mass loadings of $\eta\sim10$.  This,
however, is the exact same situation which produces bubbles with short cooling
times. Only at lower metallicities would these more mass-loaded bubbles be able
to spend an appreciable time in the CGM.

This of course assumes that the only material to be ejected at breakout is the
hot, well-mixed interior of the superbubble.  This is a rather crude
approximation, as simulations of superbubbles have shown their breakout to be a
turbulent, chaotic process involving multiphase gas spanning orders of magnitude
in density and temperature \citep{Joung2006,Kim2018,Li2017b,Fielding2018,El-Badry2019}.
Treating this self-consistently is intractable for a simple analytic framework.
The results of \citet{Kim2018}, for example, show that the mass loadings of warm
gas exceeds that of hot gas when measured close to the disc, but that drops
precipitously as you move higher, with hot gas mass loadings staying constant at
$\eta\sim0.1$, and that the interactions between the hot and warm/cool
phases of the outflow are more complicated than either total entrainment or
complete decoupling.  We can, however, make some qualitative predictions.
First, that the hot, unmixed bubble component provides a floor for the mass
loading, as mixed or entrained material will only raise the total amount of
material lofted out.  This material will increase the mass loading, but also
decrease the acceleration of the bubble, reducing the maximum height it reaches.
Whether entrained material will cool the bubble will depend sensitively on the
quantitity and metallicity of the entrained material, and whether that material
stays as cool clouds or mixes efficiently with the bubble.  Simulations of cool
clouds in fast $v>>100\kms$ winds find that they do mix quickly
\citep{Scannapieco2015}, but this can be slowed presence of magnetic fields and
cosmic rays \citep{Girichidis2018a} or stopped through cloud ``shattering''
\citep{McCourt2015,Gronke2019}.  None of these studies, however, have looked at
the long-term kinematics and survival of multiphase material in a slowly rising,
buoyant outflow.  We plan to examine simulations of this scenario in a future
study.

Radiative cooling isn't the only potential instability that may disrupt a bubble
as it rises.  The bubble itself can be seen as the leading finger of a
Rayleigh-Taylor instability, and will thus generate vorticity in it's trail,
potentially mixing it as it takes the classic ``mushroom cloud'' shape of a
Rayleigh Taylor plume.  The edges of the bubble, as well, will experience shear
against the background CGM, subjecting them to Kelvin-Helmholtz instabilities.
We know from X-ray observations of galaxy clusters that, at least in the extreme
AGN-driven case, bubbles can rise to a significant fraction of the cluster
virial radius (see for example the $\sim200\kpc$ bubbles observed in
MS0735+7421, \citealt{McNamara2005}) before being completely mixed with the
background intercluster medium.  Simulations of either AGN-driven
\citep{Quilis2001,Omma2004,Bruggen2009,Dong2009} or SN-driven \citep{Sarkar2015}
have found that wind-driven bubbles do experience both classic RT-like
instabilities, as well as more complex non-linear turbulent mixing.
Interestingly, the simulations of SN-driven starbursts presented in
\citep{Sarkar2015} found that the mixing in the CGM can actually increase the
mass loading near the virial radius by up to a factor of 10-40 by sweeping up
CGM material.  The Kelvin-Helmholtz instabilitiy was analytically shown to be
supressed in the presence of viscosity or in inviscid bubbles when there is a
significant difference in the mass density of the bubble and the light fluid by
\citep{Kaiser2005}.  They find that for Rayleigh-Taylor instabilities, only
viscous effects producing an effective surface tension can preserve the bubble.
Even with a conduction coefficient 100 times below the standard Spitzer value, a
magnetic field of a few $\mu G$ can stabilize the bubble.  We would also expect
to see much longer Rayleigh-Taylor timescales in the shallower potential well of
an $L*$ galaxy (compared to the $\sim10^{14}\Msun$ halo these studies examined),
as the growth time scales linearly with the Keplerian velocity.  However, as
\citep{Dong2009} showed, the relatively high magnetic field strengths that do
stabilize bubbles in 3D hydrodynamic simulations require specific geometries,
and are generally higher than what is observed in the ICM of clusters such as
Virgo.  

These hydrodynamic instabilities mean that our treatment of the bubble as an
arbitrarily long-lived object, subject only to radiative cooling is at best a
rough approximation.  Long-term oscillations about the buoyant equilibrium
radius, as seen in figure~\ref{buoyant_flight} will not occur if the bubble is
mixed into the CGM.  If a bubble were to fragment as it rises, this will
naturally increase the overall surface area of the high-entropy gas, increasing
the drag forces it would experience.  Despite this, there are two reasons we do
not expect that bubble fragmentation will completely halt an entropy-driven
outflow. First is that the drag forces experienced by a bubble are most
significant during the very early evolution of the bubble, when it is
accelerated to the highest velocities and it travels through the densest part
of the CGM.  If the timescale for a bubble to fragment is longer than the time
it takes to rise above the dense inner CGM, drag forces become unimportant
regardless of surface area at the lower velocities the bubble spends most of
it's lifetime at.  Second, if a bubble fully mixes with the CGM, it may
actually stabilize the CGM against radiative cooling, increasing the entropy of
region it mixes into if it mixes below the buoyant equilibrium radius or by
driving turbulence in the CGM (acting thus as a kind of preventative feedback).
Simulations by \citep{DallaVecchia2004,Bruggen2009,Dong2009} all show that
while mixing is significant as an AGN-heated bubble rises, it will still reach
significant $>100\kpc$ altitudes prior to being fully mixed with the
surrounding medium.  When a buoyant bubble mixes with the CGM, as long the
subsequent material is sufficiently distant from the galaxy disc that it will
not rapidly cool and re-accrete, it will still act to transport mass and metals
out of the ISM.  In certain circumstances, however, the opposite effect may
take place, and bubbles driven out of the ISM may actually seed radiative
cooling \citep{Marasco2012} or ``precipitation'' from the CGM
\citep{Voit2017,Voit2019}, increasing the accretion rate onto the galaxy.

The uncertainty around how multiphase gas arises in galactic outflows (whether
it be through entrainment, in-situ cooling, or some other mode) is tempered by
both observational and (albeit low-resolution) simulation evidence.  In the CGM,
hot ($T>10^6\K$), and warm-hot ($T\sim10^{5.5}\K$) gas has been observed in a
wide variety of environments \citep{Werk2016,Stocke2013}, but so has cool
$T\sim10^4\K$ \citep{Tumlinson2013}, or even molecular gas \citep{Leroy2015}.
Meanwhile, simulations of $L*$ and larger galaxies have found significant mass
loadings \citep{Keller2015,Muratov2015,Christensen2016,Mitchell2019}, with much
of this mass being carried by cool material.  Bridging the gap between
high-resolution, small scale simulations and the observational/simulation
evidence on larger scales that suggest a ubiquitous, multiphase ISM will require
serious work both on the analytic, theoretical front as well as in more
sophisticated numerical modelling.

\section{Comparison to Simulations}
\subsection{Cosmological Simulations}
If cosmological simulations are run with adequate resolution \citep{Muratov2015}
and sub-grid feedback models \citep{Keller2014,Keller2015} to generate
ab-initio winds of hot, high-entropy gas, we should expect to should see
outflows being driven by entropy gradients, with the long recycling times and
relatively low velocities that entropy-driven wind theory predicts.

Here we compare to simulations from the McMaster Unbiased Galaxy Simulations 2
(MUGS2) sample of galaxies.  MUGS2 is a set of 18 cosmological zoom-in simulations,
focusing on the evolution of $L*$ galaxies with a range of merger histories and
spin parameters, selected without biasing to reproduce any given single object.
These galaxies were simulated in a {\sc WMAP3} $\Lambda CDM$ cosmology, with
parameters $H_0 = 73\kms \Mpc^{-1}$, $\Omega_M=0.24$, $\Omega_{bar}=0.04$,
$\Omega_\Lambda=0.76$, and $\sigma_8=0.76$.  The MUGS2 $z=0$ halo masses range
from $3.7\times10^{11}\Msun$ to $2.2\times10^{12}\Msun$, with disc masses
ranging from $1.8\times10^{10}\Msun$ to $2.7\times10^{11}\Msun$.  
The details of how the initial conditions for these galaxies were built can be
found in the first MUGS paper, \citet{Stinson2010}.  

Each of these simulations has a gas mass resolution of $M_{\rm
gas}=2.2\times10^5\Msun $, and uses a gravitational softening length of
$\epsilon=312.5\; \rm{pc}$, and a minimum SPH smoothing length set to 1/4 of
this.  We use a simple Schmidt-law star formation prescription \citep{Katz1992},
with an efficiency per freefall time of $0.05$.  Star formation can occur in gas
that with density exceeds $\sim10\hcc$, and temperature below $1.5\times10^4\K$.
A critical difference between the MUGS and MUGS2 simulations is the introduction
of the \citet{Keller2014} superbubble feedback model.  This model captures the
effects of thermal evaporation, as described in \citep{Cowie1977,Weaver1977} in
order to determine the proper amount of hot gas heated by SN feedback
\citep{MacLow1988}.  This allows the superbubble feedback model to generate 
high entropy, buoyant gas, as was described in
\citet{Keller2015,Keller2016}.

In figure~\ref{halo_entropy}, we show the entropy and density profiles for each
of the MUGS2 galaxies.  As can be seen, each can be roughly described as a power
law, with slopes consistent with an entropy profile with $\alpha=1.1$.  The
break in the density slope around $10\kpc$ is due to the edge of the thick disc
ISM contributing mass to the profile.  It can be seen here as well that the
normalization of the entropy profile increases with mass (the highest entropy
profile corresponds to the galaxy with $M_{200}=2.2\times10^{12}\Msun$, which
has a virial entropy of $\sim60\;\mathrm{keV\;cm^{-2}}$ as per
equation~\ref{virial_entropy}). 

If we take a look at the MUGS2 ``fiducial'' galaxy, g1536, we can see that the
entropy profile can be broken into two components (as was seen in
figure~\ref{entropy_story}):  a cooling, inflowing component, and hot,
outflowing component.  In figure~\ref{entropy_velocities}, we take this data and
bin by radius and entropy, weighting each bin by the mean velocity of gas in the
bin.  As can be seen in this figure, we do indeed see higher outflowing
velocities in gas with entropy above the profile given by
equations~\ref{virial_entropy} and~\ref{CGM_entropy}.  Gas below this curve is
almost exclusively inflowing, except for a small amount of material at
$r\sim20-50\kpc$.  This material may either be entrained gas, swept up by drag
on passing superbubbles, or it may actually be a signature of the convective
overshoot seen in figure~\ref{buoyant_flight}.  However, if we consider the mass
bins shown in figure~\ref{entropy_story}, we can see that this
overshoot/entrained material actually accounts for very little mass.

The key difference between entropy-driven and ballistic fountains is the
relationship between the outflow velocity and the distance it can travel from
the galaxy before re-accreting, as well as the time it takes to do so.  By
looking at the final $3\Gyr$ of evolution for our simulated galaxy, we can
examine fountain re-accretion unimpeded by either significant merger activity
(which can easily be mistaken as outflowing/fountain material as it recedes from
its pericentric pass) and also compare to the evolution predicted by
equation~\ref{acceleration_phys_full}, with a fairly constant virial mass (the
$z=0$ virial mass of g1536 is $\sim 7\times10^{12}\Msun$).  By selecting gas
particles that cross a spherical surface around the center of the halo, with
radius $0.1R_{200}$, we can select fountain particles (those going from inside
to out) and re-accreting particles (those going from outside to in).  Tying
fountain particles to their re-accretion allows us to track the total time the
spent in the fountain and compare this to the state of the gas as it is ejected.
In figure~\ref{fountain_time}, we show the velocity and entropy of gas particles
identified in fountains as a function of the fountain events re-accretion time.
We bin each fountain event based on the re-accretion time, which allows us to
look at the distribution of ejection velocities (top panel) and entropies
(bottom panel) for fountain events that are re-accreted over a range of
timescales.  As this figure shows, there is very little relationship between the
velocity and re-accretion time, with most fountain events having initial
velocities of $\sim100\kms$, and none having initial velocities above $350\kms$.
We show the predicted fountain recycling time here for different initial
velocities in a halo with the same mass as g1536.  Comparing the simulation
results to this prediction, we can see that the fountain events which re-accrete
in $>500\Myr$ are travelling far too slowly to be simply evolving ballistically.
If we instead  look at the entropy of fountain flows at their launching, we can
see that higher entropy outflows persist for longer prior to re-accretion, with
low entropy gas ($K<2{\rm keV\;cm^2}$) returning in $\sim500\Myr$.  High-entropy
gas ($K>5{\rm keV\;cm^2}$), with entropies typical of what we expect for
SN-driven superbubbles (given by equation~\ref{SB_entropy}) on the other hand
re-accretes on longer timescales, $>\Gyr$, comparable to the predicted evolution
for a bubble of different initial entropies travelling through the same halo as
is used for the ballistic lifetime predictions.

An important feature to note in figure~\ref{fountain_time} is the direction of
the offset between the median velocity/entropy and the predicted
velocity/entropy for the range of fountain lifetimes shown.  The ballistic
predictions for the fountain velocity are a lower limit for the lifetime.  In
these calculations, we have omitted the drag term used in
equation~\ref{acceleration_phys_full} to allow the largest possible recycling
times given the uncertainty of ballistically-driven winds interacting with the
CGM.  Any drag or mixing will act to decelerate the outflow, pushing the orange
curve further towards the upper left of the plot, and increasing the difference
between the simulation results and the ballistic theory prediction.  On the
other hand, our calculations for the flight of entropy driven outflows have
included the effects of drag, but omitted the effects of cooling.  Mixing,
stronger drag, or radiative cooling will sap the bubble of momentum or entropy,
causing it to re-accrete sooner.    This also will move our prediction up and to
the left, bringing it into better agreement with the simulation data. A full 3D
simulation treatment including hydrodynamic mixing and radiative cooling is left
for future work.

\subsection{Superbubble Entropy Generation in a Realistic ISM}
The derivations presented thus far make a number of simplifying assumptions
about the growth of a superbubble within the ISM, prior to its  break out and
buoyant uplift through the CGM.  We can test our simple, one-dimensional model
by comparing it to data from simulations of a more realistic, self-consistent
interstellar medium.  Recent studies have been able to investigate the break out
of galactic winds from tall, stratified slices of the ISM, achieving resolutions
of $\lesssim\pc$.  Not only do these simulations offer significantly better
resolution, down to the ability to resolve the effects of individual massive
stars, they also have included physics that is often omitted from larger
cosmological simulations, such as magnetic fields, cosmic rays, and radiation
hydrodynamics.  We can use the publically available data from these simulations
to verify our assumptions about the entropy of the hot interior of the
superbubble, prior to break out.  We use here the simulations of
\citet{Hill2018}.  These simulations are run with a wide variety of different
physical processes, all using the {\sc FLASH} code \citep{Fryxell2000}.
\footnote{The simulation outputs from \citet{Hill2018} are all publically
available, and are accessible at the American Museum of Natural History Research
Library Digital Repository \citet{Hill2018sup}}

\citet{Hill2018} presents a suite of 22 simulations of a $1\times1\times40\kpc$
xy-periodic slab of the ISM with outflow boundary conditions in the z-direction.
These simulations have varied gas surface densities, gas densities above the
disc, magnetization, resolution, and (the primary change) ISM heating rates.
Each of the simulations includes radiative cooling \citep{Joung2006} (assuming
solar metallicity) and a heating rate designed to approximate photoelectric
heating of dust grains.  This heating rate has an exponential falloff with
height above the midplane, with a scale height of $8.5\kpc$.  The initial
conditions use a fixed gravitational potential, with an initially isothermal
disc in hydrostatic equilibrium.  The feedback model used in these simulations
includes energy injection by stellar winds, type Ia SN, and type II SN.  Star
particles are created to give a fixed total SFR surface density, with randomly
chosen xy positions and an exponential distribution in height, with a scale
height of $90\pc$.  During the first $5\Myr$ of a star particle's lifespan,
stellar winds with a luminosity of $10^{36}\ergs$ per massive star
are injected, followed by either a single SN (meant to capture the effect of a
field core-collapse SN , occuring in $32\%$ of all stars), or between 7 and 40
SN (meant to capture the effect of clustered star formation, occuring in $47\%$
of all star particles).  The remaining SN are type Ia SN, distributed with a
scale height of $300\pc$.  Each supernovae deposits $10^{51}\erg$
worth of energy.

The ``fiducial'' simulations in \citet{Hill2018} each have a SN rate of
$34\Myr^{-1}\kpc^{-2}$, an ISM gas surface density of $13\Msun\kpc^{-2}$, and a
maximum refinement level of $\Delta x = 4\pc$.  A range of different
photoelectric heating rates, from $1\times10^{-25}\ergs$ to
$1.23\times10^{-24}\ergs$.  Two runs are done using a horizontal magnetic field
of $5\mu G$ at the midplane, which decreases with height to maintain constant
magnetic $\beta$.  Two runs are done with a reduced gas surface density of
$7.5\Msun\kpc^{-2}$, and two are done with a hotter, more diffuse IGM/CGM (one
with and one without magnetization).  Resolution is also varied, with $\Delta x
= 2\pc$ and $\Delta = 1\pc$.  The final additional run is a test with a reduced
SN rate of $17\Myr^{-1}\kpc^{-2}$.  We have selected a subsample of these
simulations to see if the entropy generation in a fully 3D, turbulent ISM that
includes radiative cooling, magnetic fields, and the pre-heating of SN regions
by stellar winds can generate hot gas with similar entropies to what is
predicted by equation~\ref{SB_entropy}.

The publically available data provided from \citet{Hill2018} includes a record
of where each massive star cluster is placed, as well as when it was formed and
how many SN have detonated.  We select each star that particle that has formed
between $t=160\Myr$ and the final timestep of $t=200\Myr$, and select
(mass-weighted) gas within a $20\pc$ sphere around these particles in this final
output.  This allows us to examine regions where SN have detonated, but have not
yet had a chance to vent out of the ISM.  We also select a smaller subsample of
star particles that have formed between $t=195\Myr$ and $t=200\Myr$ to examine
the density of the ambient ISM which SN are detonated in.  Naturally, these
small, young superbubbles may not have the same entropy at breakout as they do
when they may break out of the ISM.  Subsequent mixing of cool gas or radiative
cooling may act to reduce the entropy of this material before it breaks out.
Therefore, the values measured here should be seen as upper limits to the
entropy at breakout.  Tracking material through the halo is also, unfortunately,
impossible to do with simulations at such high resolution.  The resolution that
allows us to treat the ISM's multiphase structure in detail also limits us to
studying only small slices of the disc in a stratified, periodic box.

We use these data to examine three predictions made by our simple 1-dimensional
calculations: that sufficient entropy is generated inside superbubbles by SN to
make the superbubbles strongly buoyant, that these superbubbles have
sufficiently long cooling times to actually escape the ISM and expand
adiabatically, and that the superbubble's entropy is only weakly dependent on
the star cluster's driving luminosity.  As can be seen in the entropy Kernel
Density Estimate (KDE), figure~\ref{Hill_entropy}, regardless of the wide variety
of parameters and physics included in the different \citet{Hill2018}
simulations, we find the majority of superbubbles produce gas with a entropy
$>10\;{\rm keV\;cm}^2$, with a small number superbubbles ``failing'', due to
their detonation in particularly dense environments with short cooling times.
In every case, the median entropy of superbubbles in the simulation is above the
normalization in equation~\ref{SB_entropy}.  The magnetized case appears to
produce the lowest entropy superbubbles, while the case with lower gas surface
density produces the highest median entropy.  The variation in the median
entropy induced by varying the photoelectric heating rate is fairly small: it
appears that a higher heating rate produces a broader distribution of
superbubble entropies, but little change in the median value of this
distribution.

If we look specifically at the environment around stars which have not yet
detonated any SN, we can see in figure~\ref{Hill_density} that the majority of
SN detonate within gas with an ambient density of $\sim10^{-3}\hcc$.  This is
important both to increase the entropy of the superbubble, and to reduce the
cooling time.  As we see in figure~\ref{Hill_cooltime}, essentially all
superbubbles with $K_{SB}>10\;{\rm keV\;cm}^2$ have cooling times $>100\Myr$,
and those which would be positively buoyant all the way to a virial radius in a
$10^{12}\Msun$ halo have cooling times $>\Gyr$.  These cooling times will
increase as well, as the bubble rises through the CGM and becomes more diffuse
(as we saw in figure~\ref{cooltime_Mvir}).  The range of SN detonated by each
star particle in the \citet{Hill2018} simulations also lets us probe the
dependence of superbubble entropy on the SN luminosity of the star cluster.  A
cluster with $N_{SN}$ detonated over a period of $40\Myr$ will have a luminosity
$L_{SB} = 7.93\times10^{35}N_{SN} \;{\rm erg/s}$.  We can bin all of the star
particles in the \citet{Hill2018} by the number of SN detonated to see, as we
show in figure~\ref{Hill_clustering}, to examine whether there is any
significant trend in the entropy generated versus the cluster luminosity.  As we
can see, there is a fairly large scatter in the superbubble entropy, owing to
the inhomogeneous ISM and the different ages of the superbubbles.  The median
entropy of the full sample of superbubbles is $32\;{\rm keV\;cm}^2$, and as is
clear, there isn't an obvious trend with increasing cluster luminosity.  In the
range of luminosities we see here, assuming identical ambient ISM conditions, we
should only see an entropy increase of $5\%$ between the smallest ($N_{SN}=7$)
and largest ($N_{SN}=40$) clusters.  This is much smaller than the scatter we
see here, and thus it is unsurprising that no trend is noticeable.  The small
IQR seen in the final bin is simply a function of a small number of clusters
actually falling within that bin.  It is important to note, though, that the
\citet{Hill2018} simulations do not use a self-consistent star formation recipe,
but instead randomly place star clusters within $90\pc$ of the disc mid plane.
This will certainly somewhat over-estimate the number of SNe detonating in
low-density environments. Future studies looking at the driving of
superbubbles, with self-consistent ISM and disc structure will be able to
quantify this in much greater detail.

\section{Comparison to Observations}
The HST COS instrument \citep{Green2012} has produced a number of unprecedented
surveys of the CGM of $L*$ galaxies.  By probing absorption of background QSO
spectra by the resonant {\sc Ovi} doublet at $T\sim10^{5.2-5.7}\K$
\citep{Tumlinson2011}, it is able to detect out-of-equilibrium warm-hot gas
without relying on either high column densities needed to see neutral/molecular
clouds, or the higher temperatures needed to detect CGM gas through X-ray
emissions.  In the last 5 years, a number of large $N>10$ surveys of absorption
systems surrounding $\sim L*$ galaxies have been produced that constrain the
mass, structure, and kinematics of these galaxies' CGM
\citep{Stocke2013,Tumlinson2013,Werk2014,Werk2016}.  These studies probe samples
of galaxies at different impact factors (angular separation between the galaxy
and QSO, a lower limit on the radial distance between the galaxy and the
absorber) at fairly low redshift $z\sim0.2$.  \citet{Werk2016} presented an
interesting reduction of the COS-Halos data, selecting in particular galaxies
with active star formation, relative isolation, and positive detection of {\sc
Ovi} within their inner CGM.  These are a ``cleaner'' subsample of the full
COS-Halos dataset as they are least likely to be contaminated by material
stripped from interacting haloes.

In figure~\ref{temp_Mvir}, we show the temperature evolution of a bubble as it
rises through the CGM, while remaining in pressure equilibrium with the
surrounding medium.  As it rises and expands, the bubble's initial temperature
decreases adiabatically.  Naturally, bubbles driven by larger clusters have
higher initial temperatures at breakout, but significantly less than the 2
orders-of-magnitude difference in mechanical luminosity.  As can be seen in the
green band, the bubbles pass through the temperature range where {\sc Ovi}
absorbtion would be strong ($f_{OVI}>0.02$)for a few $10-100\Myr$, and for
bubbles driven by larger clusters, will be detectable in the CGM as they
oscillate back down to higher CGM pressures, when their temperatures increase
from adiabatic compression.  While this timeframe is a relatively small fraction
of the time the bubble spends in the CGM, a star forming disc, producing
multiple star clusters continuously driving feedback, will populate the the CGM
with multiple bubbles at different heights, giving a detectable signature
throughout the halo.  It is also possible that bubble material may mix with
hotter material in the CGM, cooling it to the temperatures where {\sc Ovi} will
be detectable.  During the bubble's initial flight, when it's temperature is
above the narrow range where {\sc Ovi} abundance is high, entrained cool
material from the ISM would also produce {\sc Ovi} absorbtion features in the
mixing layer between these cooler clumps and the hot bubble material.

Here we compare our predicted radius-velocity phase-space behaviour to the
observed data from two of these studies, \citet{Stocke2013} and
\citet{Tumlinson2013}.  Both of these focused on galaxies near $L_*$, and probed
the kinematics of CGM absorbers out to a few $R_{200}$.  As we see in
figure~\ref{velocity_radius}, none of the observed CGM absorbers found in these
COS observations exceed the escape velocity of a $\sim L_*$ galaxy, with
$M_{halo}=7\times10^{12}\Msun$ and $v_{esc}\sim700\kms$.  There also appears to
be little relation between the impact factor of the observation and the velocity
centroid, with velocities of a few to a few $100\kms$ seen from $\sim10\kpc$ out
to $\sim300\kpc$, beyond the virial radius of most of these galaxies.  We show
contours of the predicted phase-space behaviour for entropy driven outflows for
a $7\times10^{11}\Msun$ halo being driven by clusters with luminosity (mass) of
$10^{38}\ergs$, $10^{39}\ergs$, $10^{40}\ergs$ ($10^{4}\Msun$, $10^{5}\Msun$,
$10^{6}\Msun$).  As these contours show, entropy-driven outflows can produce
outflowing material at the same radii, with the same velocities as the warm {\sc
Ovi} absorbers observed by COS.

COS-Halos selected galaxies in part based on their SDSS $k-$corrected masses
derived from $ugriz$ magnitudes, giving a dataset which includes estimates of
the total stellar mass and SFR for their sample of galaxies.  Using these
stellar masses, we can use a stellar mass to halo mass relation (SMHMR)
determined either through abundance matching \citep{Behroozi2013} or directly
with weak lensing \citep{Hudson2015}.  Here, we use equation C1 from the
CFHTLenS observed SMHMR to determine halo masses for each of the galaxies in
COS-Halos, to compare with the predictions of entropy driven winds in different
mass haloes.  As we see in figure~\ref{cos_haloes},  outflowing metals are seen
with a wide range of velocities, without a strong dependence on halo mass.  We
show the predicted range of velocities as time-weighted violin plots (as the
longer gas stays at a given velocity, the higher its chance of being observed at
that velocity) for a range of different halo masses, using the same range of ISM
densities and scale heights as figures~\ref{apsis},~\ref{t_apsis},
and~\ref{velocity_radius}.  As can be seen from these violin plots,
entropy-driven winds can produce some high-velocity outflows (in the cases of
very high-luminosity clusters driving outflows from a very low density, thick
ISM) even up to $M_{200}=10^{13}\Msun$, but both the median and maximum
velocities decrease at higher halo mass.  A fit to the \citet{Werk2016}
subsample finds that the typical outflow velocity is $v_{out}\sim
30(M_{halo}/10^{12}\Msun)^{-1/3}\kms$, in good agreement with the relationship
between the median outflow velocity and halo mass for entropy-driven winds from
a wide variety of ISM environments.  It is also clear that both the observed CGM
absorbers from \citet{Werk2016} and the majority of entropy-driven outflows have
velocities well below the virial velocity of their parent halo, meaning this gas
is bound to the galaxy halo.

\section{Discussion}
We have examined here the behaviour of a new kind of galactic outflow, driven
not by feedback momentum injection in the ISM, but instead through a steady,
continuous acceleration through the CGM.  This acceleration is a consequence of
buoyant force between hot, high entropy gas and the entropy-stratified CGM.  The
kinematics of entropy-driven winds are able to broadly reproduce both
observations of gas flows in the CGM as well as the behaviour of galactic
outflows in cosmological simulations of $L*$ galaxies.

\subsection{Ballistic vs. Buoyant Winds}
What we have shown through this paper is that the phenomenology of galactic
outflows depends heavily on whether one considers the CGM environment through
which they travel.  Estimates of outflow recycling times, the mass loading of
outflows, and the distance which these outflows might reach can be off by an
order of magnitude or more if made assuming simple ballistic flight.  Simply
focusing on the energy budget available from feedback processes or the initial
velocity of feedback-heated gas heavily underestimates the effectiveness of
outflows at removing significant amounts of gas from the ISM for long,
cosmological timescales.

Taking buoyancy into account, and modelling outflows as entropy-driven, rather
than simply ballistic, solves a number of issues in galaxy evolution.   Outflows
which remove a significant fraction of material from the ISM, can  solve both
issues with the baryon \citep{Papastergis2012}, stellar \citep{Behroozi2013},
and metal \citep{Peeples2014} content observed in galaxies.  Ballistic outflows
recycle on a very short timescale, with $t_{rec}\sim100\Myr$ for $L*$ galaxies
with halo masses $\sim10^{12}\Msun$.  This means that ballistic winds would need
to remove material from the ISM $10-100$ times to prevent it from forming stars,
compared to the $1-10$ times required by entropy-driven winds. On top of this,
these fountains can rarely exceed apocentric distances of a few $10\kpc$,
rendering it hard to see how they are able to populate the CGM with metals and
cool material out to the $\sim R_{200}$ distances they have been observed
\citep{Stocke2013}.  Slow-moving winds, with $v\sim100\kms$ out to
$r\sim100\kpc$ are a natural consequence of allowing outflowing gas to be
accelerated through the CGM.

\subsection{Comparison with Other Models}
The idea of buoyancy as critical feature of SN-driven outflows in $L*$ galaxies
was first explored by \citet{Bower2017}.  \citet{Bower2017} argues that the
transition in galaxy properties seen at $M_{200}\sim10^{12}\Msun$ is a result of
the halo's virial entropy exceeding the characteristic entropy of
supernovae-heated gas.  This is similar to what we have explored here, with a
few critical differences.  \citet{Bower2017} is not a full exploration of the
kinematics of entropy-driven outflows, but instead uses a number of simple
scaling relations, to derive a characteristic entropy for supernovae-heated gas,
with a different functional form to what we have derived here.  Unlike our
derivations, which follow the evolution of a wind-swept bubble powered by a
single star cluster, the \citet{Bower2017} simply parameterize the mass-loading
and assume that the ISM density and halo mass follow a simple scaling relation,
with $K_{outflow}\propto\eta^{-1}$.  While their derivations produce slightly
different halo entropy to what we have used here, it is roughly comparable to
what we derive in equation~\ref{halo_entropy}.  \citet{Bower2017} uses their
parameterized feedback entropy to suggest that the outflow mass loading $\eta$
can only be large when it allows $K_{outflow} > K_{halo}$, which limits this to
haloes with $M_{200} < 10^{12}$.  What we have derived here is a self-consistent
model of how the ISM structure and the star formation process sets the entropy
of outflowing gas (for most reasonable values of $n_{amb}$, $L_{38}$, and $h$,
we find entropies within a factor of a few compared to what \citealt{Bower2017}
finds), and how this gas actually moves through the CGM.  Our results are in
broad agreement with the central result of \citet{Bower2017}, namely that above
$M_{200} = 10^{12}\Msun$ supernovae-driven outflows fail to effectively remove
material far beyond the disc of galaxies in $M_{200}\sim10^{12}\Msun$ haloes (as
is shown in figure~\ref{apsis_cool} and~\ref{cos_haloes}).  However, the
simplified assumption that effective outflows require $K_{outflow} > K_{halo}$
ignores two of the key results we have presented here. First, that the CGM's
entropy stratification can allow entropy-driven winds to reach a significant
fraction of $R_{200}$ without actually leaving the halo. Second, outflows with
entropy below the virial entropy will overshoot their equilibrium radius due to
simple momentum conservation, and can take a significant amount of time to
either reach their maximum height or to cool and re-accrete.  

A recent study \citep{Lochhaas2018} has produced similar results to what we have
shown here, using a quite different analytic framework.  Like this work,
\citet{Lochhaas2018} builds on the classic wind-driven shell solutions derived
by \citet{Castor1975,Weaver1977} to predict the kinematics of gas flows within
the CGM.  Unlike this work, however, \citet{Lochhaas2018} builds off an
assumption that the galaxy can be treated as a single luminosity source, with
superbubbles escaping from the ISM and combining to form a uniform, spherically
symmetric wind-swept bubble.  As this shell propagates outward, it sweeps up
material from the CGM to build up a fairly slow-moving, massive shell.  Despite
having a relatively low velocity the wind-swept shell can take $>\Gyr$ to
re-accrete, both because the initial wind velocity is high ($\sim400-900\kms$)
and the shell itself is continuously receiving momentum from the wind behind it.
\citet{Lochhaas2018} shows that for the majority of the time it spends in the CGM, it
travels with modest $\sim100\kms$ velocities, and that these kinematics and the
column density profile of the shells are consistent with the COS-Halos
observations of \citet{Werk2014}. Our work differs from this analysis primarily
in that we begin with a local treatment of a superbubble driven by a single star
cluster within the ISM.  Our model does not require a continuous,
globally-uniform starburst to drive outflows.  One of key feature of
our model is that it couples the local environment of star formation to the
global effects of outflows and fountain recycling.  Entropy-driven winds do not
require a uniform shell to be swept up, but can rise as individual distinct
bubbles through a static background CGM.  Future observations may be able to
distinguish between these two pictures \citep{Spilker2018}.  Isolated,
entropy-driven bubbles would produce a velocity distribution through the ISM
with much less radial isotropy than would be produced by a global starburst.

\subsection{Is the Cooling CGM Bubbly or Windy?}
We have seen clearly that the stability of outflowing gas to radiative cooling
will depend heavily on the mass loading in the outflow.  The mass loading
directly sets the initial temperature of the rising bubble, as the average
bubble temperature is $T\sim (5.67\times10^7\K)\eta^{-1}$, and thus the bubble's
susceptibility to cooling.  As was explored in \citep{Thompson2016}, a global
galactic outflow powered by disc-wide SN will produce a wind roughly following
the classic \citet{Chevalier1985} solution.  The cooling radius of this wind,
for mass loadings $\eta>0.5$, is typically $\ll100\kpc$, and the cooling times
are typically much less than a Hubble time.  This cooling may result in a
shell-like structure, where a cool shell can be accelerated and compressed out
of the CGM, and may actually persist for $>\Gyr$ despite this cooling, as was
explored recently in \citep{Lochhaas2018}.  Alternatively, this may not drive a
stable, roughly spherically-symmetric set of cool shells, but instead fragment
to form pressure-confined cool clouds embedded in a hot, diffuse medium
\citep{Maller2004}, driven by \citet{Vishniac1983} or similar
hydrodynamic instabilities in the shocked wind.  A critical difference between
our approach and these previous ones is the treatment how galactic winds
develop.  Here, we have assumed that the fundamental unit for treating outflows
are bubbles driven by individual star clusters, while
\citep{Chevalier1985,Thompson2016,Lochhaas2018} assume that disc-wide star
formation drives a coherent, spherically symmetric wind through the CGM.  An
interesting question for a future study is how vigorous star formation must
become for bubbles driven by individual star clusters for those bubbles to merge
and produce a global, uniform wind.  Is the CGM best characterized as a global,
spherically symmetric \citep{Chevalier1985} style wind, or instead a series of
convective bubbles rising and falling as high-entropy bubbles are injected and
low-entropy clumps cool out?  The answer will have significant consequences on
the observational characteristics of the CGM absorption/emission features, as
well as the overall energy budget required to maintain a quasi-stable atmosphere
in $L*$ galaxies.

\subsection{Everything Counts (in Some Amount)}
The model we have presented here couples together the state of four major
components of the galaxy: the ISM, the CGM, the stellar population, and the DM
halo.  Each of these components play a role, either in setting the initial
entropy of feedback-heated gas (the density and scale height of the ISM, and
the mass function of young stellar clusters), the acceleration of gas through
the CGM (the entropy profile of the CGM), or the deceleration through drag or
gravity (the density profile of the CGM and the DM halo).  This gives us the
opportunity to use this model to examine the effects of each of these on each
other, and on the evolution of the galaxy as a whole.  

We have interpreted the relations derived in equation~\ref{SB_entropy} in terms
of SN driving by a stellar cluster, but these equations are valid for any
continuous injection of energy, and would lend themselves naturally to heating
by AGN as well.  The much higher driving luminosity of AGN ($>10^{44}\ergs$
\citealt{Tremmel2017}), as well as their potential to heat gas beyond the disc
of the galaxy \citep{Sijacki2007} means that they will be able to generate far
greater entropies than SN-driven superbubbles, and produce gas that is buoyant
even in cluster-mass ($\sim10^{14}\Msun$) galaxy haloes.  Indeed, buoyancy is
already used as a mechanism for understanding the flight of X-ray bubbles
through the hot coronae of galaxy clusters \citep{Sanders2005,Voit2017}.

The very weak dependence on driving luminosity that we derive in
equation~\ref{SB_entropy} ($K_{SB}\propto L_{38}^{2/63}$) means that even for
the largest observed star clusters ($M\sim10^7\Msun$), the entropy of gas within
that cluster's superbubble will be only $\sim30\%$ greater than that of a
cluster with only 10 massive stars $M\sim10^3\Msun$.  This helps in part to
explain why stellar feedback becomes ineffective above
$M_{halo}\sim10^{12}\Msun$, and why AGN feedback may be required to effectively
regulate the star formation and baryon content of these galaxies
\citep{Keller2016,Bower2017}.

\subsection{Caveats and Future Work}
While we have shown that this model can reproduce both observations and
simulations of the CGM, and that our assumptions about the generation of entropy
by feedback in the ISM are broadly in agreement with high-resolution ISM
simulations, we have made a number of simplifying assumptions throughout this
work.


The primary simplification we have made in developing the equations governing
entropy-driven winds is the simple 1D approximation used throughout section 2.
We have assumed here that both the growth of a superbubble in the ISM as well as
the structure of the CGM can be approximated as radially symmetric.
Furthermore, we have assumed that the of the CGM is both well-approximated by a
power-law stratified entropy profile, and is in rough hydrostatic equilibrium.
It is known from theoretical arguments \citep{Nulsen1986} and simulations
\citep{Dekel2006} that galaxies are fed by cool, denser streams for much of
their life, especially at high redshift $z>1$.  How entropy-driven winds may
interact with these filaments will require further study.  We have also modelled
the mass profile of the galaxy as a singular isothermal sphere, with a rotation
speed set only by the virial mass $M_{200}$.  Further work will be needed to
examine the behaviour of entropy-driven winds in different CGM profiles, as well
as to examine the effects of inhomogeneous, non-radially symmetric CGM
atmospheres.

In addition to the simplifications made in the geometry of the galaxy halo, we
have opted here to use a very simple analytic model for the high entropy
superbubble.  We do not use the profile in $\rho$ and $T$ for the interior of
the bubble derived by \citet{Weaver1977} and \citet{MacLow1988}, but instead
assume that the bubble becomes homogenized during break out and the subsequent
rise through the CGM.  Treating the internal structure of the bubble in a more
self-consistent way would yield a spectrum of recycling/cooling times for
feedback heated gas, with the hottest, most diffuse gas inside the superbubble
travelling the furthest from the galaxy and recycling/cooling on the longest
timescale (which would in turn set where in the CGM dust and metal-rich gas is
deposited).  Determining the degree to which this might result in shear-driven
mixing inside the bubble, driving it towards the simple homogeneous state we
assume here will require high resolution 3D simulations of both superbubble
break out and its subsequent flight through the CGM.  This could easily be
combined with a self-consistent treatment of the cooling within this bubble,
which would allow the loss of entropy through radiation to couple to the
acceleration due to buoyancy.  Recent simulations are beginning to reach
sufficient resolution $(\sim\pc)$ to resolve this process
\citep{Kim2018,Schneider2018}.

\section{Conclusion}
In this paper, we have presented a novel theory for understanding the mechanism
for accelerating outflows and fountains from galaxies, driven through
hydrodynamic interactions between feedback-heated gas and the CGM.  These
entropy-driven winds have modest characteristic velocities ($\sim100\kms$), well
below the escape velocities of their parent haloes. Despite this, these outflows
can spend a significant time in the CGM, depleting the galactic disc of metals
and baryons, and slowing star formation by removing its fuel.  As we find, the
results of both cosmological simulations of $L*$ galaxy evolution and
high-resolution simulations of superbubble growth in the ISM are in broad
agreement with the predictions of this model.  We also find that these
entropy-driven winds well-reproduce the key observations of the CGM kinematics
probed by COS-Halos \citep{Werk2014}.

Entropy-driven winds may be the key to producing galaxies with low baryon and
stellar fractions, and to distributing the metals produced by star formation far
from the disc of galaxies with $M_{200}\sim10^{12}$ or below.  Alternative
mechanisms for launching outflows, especially simple ballistic launching after a
brief, sharp injection of momentum by stellar feedback cannot produce highly
mass loaded winds with sufficient velocity to reach high galactocentric radii,
or to avoid re-accreting in $<<\Gyr$, even if hydrodynamic drag and mixing is
ignored.  Entropy-driven winds on the other hand, can easily reach $>100\kpc$
above the disc, and re-accrete in a significant fraction of a Hubble time.  In
more massive haloes, we have confirmed that the increased CGM entropy does
suppress these outflows, as was predicted in \citet{Bower2017}.

Entropy-driven winds are not new physics.  They rely on simple, classical
hydrodynamics \citep{Chandrasekhar1961,Voit2017} and the self-similar evolution
of superbubbles \citep{Weaver1977,MacLow1988}.  By taking into account the
generation of entropy by feedback, and the buoyancy of feedback heated gas, we
have now developed a model for the galactic baryon cycle that shows how feedback
from stars, the structure of the ISM, and the stratification of the CGM
couple together to determine the kinematics and timescales involved in fountain
flows between the ISM and CGM.  We look forward to seeing this model applied to
understand the evolution of simulations and results of observations, and
incorporated into future (semi-)analytic models of galaxy formation and
evolution.

\section*{Acknowledgements}
The analysis was performed using yt (\texttt{http://yt-project.org},
\citealt{yt}) and pynbody (\texttt{http://pynbody.github.io/},
\citealt{pynbody}). We thank G. Mark Voit, Jessica Werk, and S. Peng Oh for
valuable conversations regarding this paper. We especially would like to thank
Jessica Werk for providing data from COS-Halos and Mordecai-Mark Mac Low for
pointing us towards the \citet{Hill2018} simulations, and for the authors making
their outputs available through the American Museum of Natural History Research
Library Digital Repository.  We also would like to thank the anonymous referee
for the helpful suggestions on improving this paper. The simulations were
performed on the clusters hosted on \textsc{scinet}, part of ComputeCanada. We
greatly appreciate the contributions of these computing allocations. We also
thank NSERC for funding supporting this research. BWK and JMDK gratefully
acknowledge funding from the European Research Council (ERC) under the European
Union's Horizon 2020 research and innovation programme via the ERC Starting
Grant MUSTANG (grant agreement number 714907).  BWK acknowledges funding in the
form of a Postdoctoral Research Fellowship from the Alexander von Humboldt
Stiftung. JMDK acknowledges funding from the German Research Foundation (DFG) in
the form of an Emmy Noether Research Group (grant number KR4801/1-1).
\bibliographystyle{mnras}
\bibliography{references}

\end{document}